\documentclass[%
amsmath,amssymb,
 aps,
 reprint,
prapplied
]{revtex4-2}

\usepackage{amsmath}
\usepackage{amssymb}
\usepackage{mathtools}
\usepackage{graphicx}
\usepackage{dcolumn}
\usepackage{bm}
\usepackage{xcolor}
\usepackage{ulem}

\DeclareUnicodeCharacter{2009}{\,} 
\usepackage{hyperref}
\begin{document}

\preprint{APS/123-QED}

\title{Spoofing resilience for simple-detection quantum illumination LIDAR}

\author{Richard J. Murchie}
\affiliation{Advanced Concepts Team, European Space Agency, Keplerlaan 1, 2201 AZ, Noordwijk, The Netherlands.}
\author{John Jeffers}
\affiliation{Department of Physics and SUPA, University of Strathclyde, John Anderson Building, 107 Rottenrow, Glasgow G4 0NG, United Kingdom}

\date{\today}

\begin{abstract}
Object detection and range finding using a weak light source is vulnerable to jamming and spoofing attacks by an intruder. Quantum illumination with nonsimultaneous, phase-insensitive coincidence measurements can provide jamming resilience compared to identical measurements for classical illumination. We extend an experimentally-feasible object detection and range finding quantum illumination-based protocol to include spoofing resilience. This approach allows the system to be characterised by its experimental parameters and quantum states, rather than just its detector data. Therefore we can scope the parameter-space which provides some spoofing resilience without relying upon the prohibitive method of acquiring detector data for all combinations of the experimental parameters. We demonstrate that in certain regimes the intruder has an optimal relative detection basis angle to minimise the induced error. We also show that there are spoofing-vulnerable regimes where excessive background noise prevents any induced error, while it is still possible to perform object detection, i.e. our detectors have not been fully blinded. The sensing protocol which we describe can allow for the recognition of intrusion and the possible detection of our trustworthy return signal. Our results reinforce that quantum illumination is advantageous for spoofing resilience compared to a classical illumination-based protocol.
\end{abstract}

\maketitle
\section{Introduction}
Passive remote-sensing is fundamentally insecure as it relies upon uncontrolled incident light scattered from an object to form the image. An adversary can deploy countermeasures to attempt to hide this object by jamming (denial of sensing by excessive noise injection) or spoofing (attempting to fool the sensor operator into believing false information). An active remote sensing system, such as LIDAR, is more secure than passive remote sensing as control of the light beam provides more knowledge about the expected return signal imprinted by the presence and distance of the possible target object. Nevertheless, an active remote sensing system can still suffer from jamming and spoofing attempts that it would be useful to mitigate against. Active remote sensing is not used in some situations because the extra light directed to the target object can be seen, again allowing countermeasures to the sensing to be deployed. One possible solution to this is to extract the most information from the minimum amount of active sensing light, which implies exploiting its quantum properties.

Much theoretical and experimental work has been done on object detection and range finding protocols which use twin beam quantum illumination (QI) to improve performance in comparison to a suitable classical benchmark \cite{Lloyd08,Tan08,Nair20,Guha09,Barzanjeh15,Zhuang17,Lopaeva13,Zhao22,England19,Liu20,Yang21,Yang22,Frick20,Kuniyil22,Gregory20,Gregory21,Murchie24}. It has been shown that simple-detection QI (that uses Geiger-mode click/no-click photodetectors) can provide jamming resilience from stationary or dynamic background noise when undertaking object detection and range finding \cite{Mrozowski24}. The strong non-classical inter-beam photon number correlations of the light allow simple-detection QI to facilitate effective noise-filtering.

Various quantum-secure imaging (QSI) protocols with spoofing resilience capabilities have been proposed and implemented \cite{Malik12,Roga16,Yao18,Zhao19,Heo23,Jennewein23,Wang24,Heo24,Johnson24,Espinoza24}. Moreover, in Ref~\cite{YWang25} they consider the security of a quantum range finding protocol where the target object has an active role in verifying their position. The security of these protocols is enhanced by the laws of quantum mechanics and they are in effect modified quantum cryptography schemes \cite{Bennett84}. Initial work used a modified BB84-inspired prepare-and-measure scheme for imaging, whereby the transmitter (conventionally Alice) randomly chooses a basis for which there are two quantum states to randomly choose from to use as the signal \cite{Malik12}. The receiver (conventionally Bob) measures this light in the same basis chosen by Alice. As Alice and Bob are co-located, we refer to them both as Alice henceforth. This co-location means an insecure public channel is not required and it is easy to ensure that the measurement is in the correct basis. The intruder (conventionally known as Eve) attempts spoofing via an intercept-resend strategy: she (partially or fully) intercepts the signal, measures it and then resends it according to how she wishes to spoof Alice \cite{Bennett92}. For example, Eve can attempt to spoof target object spatial or range information by sending light towards Alice in a different spatial mode or time-bin, respectively. The security of this system is derived from the information advantage (the choice of basis) that Alice enjoys. This information advantage ensures that Eve introduces errors on average, as she is unable to know perfectly the quantum state being sent from just a measurement; therefore, her resent quantum state may differ \cite{Huttner94}. QSI differs from quantum cryptography, as Eve's strategy is to deceive Alice rather than to obtain a supposed secret key. Here we extend the simple-detection QI-based object detection and range finding protocol from Ref.~\cite{Murchie24} to include spoofing resilience. We also build on work pertaining to non-classical correlation based quantum-secured imaging (QSI) \cite{Heo23,Heo24}. Furthermore, we choose non-classical correlation based-QSI instead of prepare-and-measure (BB84-inspired) QSI with a weak coherent-state pulse due to its noise-resilience (for improved SNR) and intrinsic timing-correlation for range finding.

The paper is organized as follows. In Section~\ref{sect:sys_overview} we describe an overview of the system. This includes details about the states of light involved and how we model the system measurements. Section~\ref{sect:prelim} gives the details about the security of this system. Section~\ref{sect:error_analysis} gives an analysis of the errors involved in this system. In Sect.~\ref{sect:recognition_eve} we explain how Alice can recognise a spoofing attack by Eve. In Sect.~\ref{sect:removal} we demonstrate how Alice can remove untrustworthy detector data due to the intrusion of Eve. Performance analysis of the different systems we consider is given in Sect.~\ref{sect:perform}. In Sect.~\ref{sect:thresholding} we give thresholds for erroneous conclusions about the detector data when restricted to limited number of samples. We present simulation results of a spoofing attack in a range finding problem in Sect.~\ref{sect:rangefinding}. We conclude this paper with a discussion of the contents of this paper, the results and the outlook for future research in Sect.~\ref{sect:discussion}.

\section{Spoofing resilience}
\subsection{System overview}
\label{sect:sys_overview}
We want to be able to detect the presence of an intruder \cite{blakely22,blakely24}, to determine the whereabouts of any trustworthy real signal (which may contain information about a target object) and to subtract any false information from detection statistics \cite{Roga16,Heo23,Heo24,Johnson24}. This section gives a theoretical analysis and overview of a system that does this. The system is based on a simple-detection object detection and range finding protocol with non-classically correlated quantum states. The quantum states have two sets of non-classically correlated modes: the idler modes and the signal modes. The idler light field is measured locally at the idler detectors and does not interact with a possible target object, whereas the signal light field is sent towards the possible target object (assuming the signal light field is in a path untampered by Eve). Reflection of this light field from the target object is recorded by Alice's signal detectors. Measurement of the idler modes conditions the signal modes due to the signal-idler correlations. These correlations can be used to enhance the sensitivity of the signal mode measurement. As the system uses simple detection the detectors can only register a click or no-click. Therefore, from our theoretical model of the system, we calculate detector click probabilities. It is these click probabilities that allow us both to recognise the presence of an intruder and to detect the presence or absence of a target object.

We assume that our source produces polarisation-correlated pairs of photons and we use this degree of freedom in our protocol. We could equally-well choose another type of encoding, one which is fully preserved on scattering from an object, but the theory described here is independent of this choice. The state of light used is written in the Fock basis as\begin{equation}\begin{split}
    \hat{\rho}&=c_0 \vert 0,0,0,0\rangle\langle 0,0,0,0\vert+ \\&+c_1\left(\vert 0,1,0,1\rangle\langle 1,0,1,0\vert+\vert 1,0,1,0\rangle\langle 0,1,0,1\vert\right),\label{eq:state}\end{split}
\end{equation} where the mode labelling is in the order (idler: horizontal, ..., signal: antidiagonal) : I:H, I:V, S:V, S:H or I:D, I:A, S:D, S:A, for a rectilinear or diagonal basis choice respectively and Alice randomly chooses whether the state is in the rectilinear or diagonal basis. We refer to the system that uses the state described in Eq.~\ref{eq:state} as the QI system and this could be produced from Type-II SPDC, for example. Also, we omit the off-diagonal parts of the density operator as the detectors are only sensitive to its diagonal elements. The coefficients are $c_0=\frac{1}{(\bar{n}+1)^2\mathcal{N}_0}$ and $c_1=\frac{\bar{n}}{(\bar{n}+1)^3\mathcal{N}_0}$, where $\mathcal{N}_0=\frac{1}{(\bar{n}+1)^2}+\frac{2\bar{n}}{(\bar{n}+1)^{3}}$ is a normalisation factor necessitated by the truncation of multi-photon contributions. From the mode labelling of Alice's state of light, we can see that the correlated polarisations for the idler and signal modes are as follows 
\begin{align} 
\mathrm{I:H}&\to\mathrm{S:V},\nonumber \\\mathrm{I:V}&\to\mathrm{S:H}\nonumber,\\ \mathrm{I:D}&\to\mathrm{S:D}\nonumber, \\ \mathrm{I:A}&\to\mathrm{S:A}\nonumber.
\end{align} 
These mode label mappings inform what we describe as correct and wrong coincidence clicks.

In order to consider the impact of a spoofing attempt we imagine that there are two channels at Alice's signal detector which we dub the real channel $\mathfrak{R}$ and the false channel $\mathfrak{F}$. These channels can be separated spatially and/or temporally, for example, if Eve wishes to spoof the spatial and/or temporal information. The real channel possibly includes Alice's light, some of Eve's light and background light, whereas the false channel possibly includes some of Eve's light and background light. These channels are what Alice measures. Appendix~\ref{appendix:parameters} describes the parameters for the system and Appendix~\ref{appendix:components} onwards derives the click probabilities for the system.

We characterise the strength of intrusion by its proportion $p$, the probability that Eve intercepts Alice's light, with $p=1$ meaning that all of Alice's light is intercepted and $p=0$ meaning that there is no interception. The proportion of intrusion $p$ can be partitioned into the intruded light resent towards the real channel $p_\mathfrak{R}$ and false channel $p_\mathfrak{F}$, where $p=p_\mathfrak{R}+p_\mathfrak{F}$. For our system we consider that Alice has a choice of two measurement bases. Following from this, Eve has a choice of which basis (out of the two options) she measures with. The fact that Eve must make a choice of basis implies that she does not know which basis Alice used, i.e. this is the information disadvantage that Eve has. Furthermore, Eve's choice of which basis could be biased or random and we signify this biasing with the parameter $r$. For example, $r=\frac{1}{2}$ signifies random basis selection, $r=0$ signifies that she only measures in one basis and $r=1$ signifies that she only measures in the other basis. There could be a set of system parameters such that a particular proportion of basis choice $r$ for Eve would minimise the induced error. Eve could optimise this scenario by biasing a proportion of basis choice if she knows the geometry of the problem and Alice's experimental conditions. Otherwise, she would just randomly guess the basis angle choice and set $r=\frac{1}{2}$. However, Alice could detect Eve's basis biasing by inspecting the detection statistics of the conjugate basis to Eve's biased basis, as this would have an elevated error in comparison to the reduced error detection statistics which has the closest match to her biased basis. Her basis-biasing strategy would instruct Alice to adjust her proportion of basis choice accordingly, thereby undoing any advantage Eve gained from biasing. An example scenario is that Alice has highly attenuated H/V detection statistics compared to D/A, if Eve was to only choose the D/A basis the total induced error would be reduced, as it is rare that Eve induces errors by measuring Alice's H/V photons in the wrong basis. However, for Alice's H/V statistics the error is increased beyond what is expected from random basis selection by Eve. As there is a viable counter-attack to Eve's basis biasing this ensures that Eve is best to randomly select a basis instead ($r=\frac{1}{2})$.

For each click probability, the particular polarisations involved signify which parameters we use, as detailed in Appendix~\ref{appendix:parameters} and Appendix~\ref{appendix:components}.
The idler detector click probability (which corresponds to polarisation $\mathcal{X}$) is $\mathrm{Pr}_\mathrm{I:\mathcal{X}}$. Where $\mathcal{X}\in\{\mathcal{X}_{+},\mathcal{X}_{\times}\},\text{ with }\mathcal{X}_{+}\in\{\mathrm{H,V}\}\text{ and } \mathcal{X}_{\times}\in\{\mathrm{D,A}\}$. For light untampered by Eve, Alice's signal detector click probability (which corresponds to polarisation $\mathcal{Z}$) conditioned by an idler click (for polarisation $\mathcal{X}$) is $\mathrm{Pr}_{\mathrm{S:\mathcal{Z}\vert I:\mathcal{X}}}$. Where $\mathcal{Z}\in\{\mathcal{Z}_{+},\mathcal{Z}_{\times}\},\text{ with }\mathcal{Z}_{+}\in\{H,V\}\text{ and } \mathcal{Z}_{\times}\in\{D,A\}$.  Whether this click probability corresponds to a correct or wrong coincidence click is determined by which $\mathcal{X}$ and $\mathcal{Z}$ is measured. A polarisation state for Eve $\mathcal{Y}$ is part of the set $\mathcal{Y}\in\{\mathcal{Y}_+,\mathcal{Y}_\times\}$, with $\mathcal{Y}_+\in\{\tilde{\mathrm{h}},\tilde{\mathrm{v}}\}$ and $\mathcal{Y}_\times\in\{\tilde{\mathrm{d}},\tilde{\mathrm{a}}\}$. Section~\ref{sect:prelim} provides more details about the polarisation basis elements for Eve, where we define $\theta_\mathrm{T}$ which relates Eve's basis elements to Alice's basis elements. Eve's signal detector click probability (which corresponds to polarisation $\mathcal{Y}$) conditioned by an idler click (for polarisation $\mathcal{X}$) is $\mathrm{Pr}_{\mathrm{S:\mathcal{Y}\vert I:\mathcal{X}}}(\theta_\mathrm{T})$. Eve does not have knowledge of the idler measurement, but the state she receives is still influenced by it. Alice's signal detector click probability when Eve's resent light is incident is $\mathrm{Pr}_\mathrm{S:\mathcal{Z}\vert E:\mathcal{Y}}(\theta_\mathrm{T})$, this is where Alice records a click at her signal detector corresponding to polarisation $\mathcal{Z}$ and Eve had sent out a single-photon with polarisation $\mathcal{Y}$. Whether Alice registers the light that Eve resent as a correct or wrong coincidence click is determined by which idler click ($\mathcal{X}$) and signal click ($\mathcal{Z}$) is measured.

\begin{center}\begin{figure}
\includegraphics{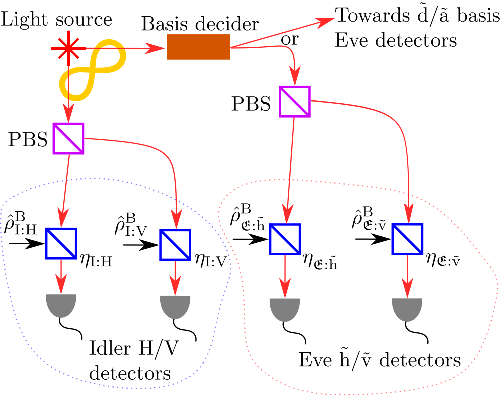}
\caption{Beamsplitter diagram of the intrusion of Eve upon Alice's signal mode light, if Alice measures in the H/V basis and Eve measures in her $\tilde{\mathrm{h}}/\tilde{\mathrm{v}}$ basis\label{fig:oursystem}. This figure visualises the beamsplitter model which we use to calculate the click probabilities for the idler detectors $\mathrm{Pr}_\mathrm{I:\mathcal{X}_+}$ and the click probabilities for Eve's detectors $\mathrm{Pr}_{\mathrm{S:\mathcal{Y}_+\vert I:\mathcal{X}_+}}(\theta_\mathrm{T})$. The beamsplitter transmission parameter relevant to Alice's idler detectors or Eve's signal detectors is set by the corresponding detector quantum efficiency $\eta_{\mathrm{I}/\mathfrak{E}}$.}
\end{figure}
\end{center}

\begin{center}\begin{figure}
\includegraphics{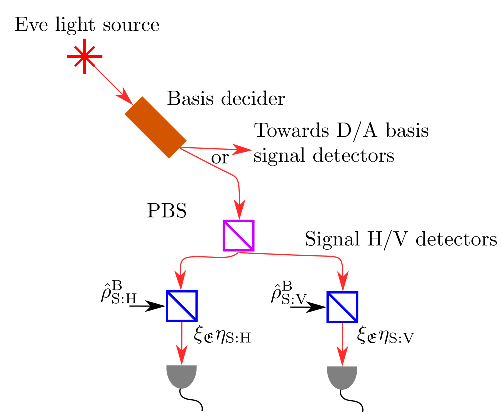}
\caption{Beamsplitter diagram of Alice's H/V signal detectors when Eve's light is incident. This choice of basis is predetermined from Alice's idler detector system not pictured. In this figure, the click probabilities for Alice's signal detectors relates to $\mathrm{Pr}_\mathrm{S:\mathcal{Z}_+\vert E:\mathcal{Y}}(\theta_\mathrm{T})$. Here, the beamsplitter transmission parameter relevant to Alice's signal detectors is the product of the detector quantum efficiency $\eta$ and signal attenuation (set by Eve) $\xi_\mathfrak{E}$.\label{fig:evesystem}}
\end{figure}
\end{center}

Figure~\ref{fig:oursystem} is a beamsplitter diagram that shows how Alice's light (when the H/V basis is chosen) is spatially separated according to polarisation and measured at the idler detectors. It also shows that Alice's signal mode light is intercepted by Eve, where she decides upon her measurement basis (whether it is the $\mathrm{\tilde{h}/\tilde{v}}$ basis as expanded upon in the figure, or the $\mathrm{\tilde{d}/\tilde{a}}$ basis). Eve then spatially separates the light field corresponding to the basis eigenstates. For the polarisation degree of freedom this spatial separation can be done with a polarising beamsplitter (PBS). Figure~\ref{fig:oursystem} does not visualise the full real/false channel system, which is what Alice actually measures. As specified by Eq.~\ref{eq:real_defn} and Eq.~\ref{eq:false_defn} the real/false channel click probabilities can comprise of contributions from Alice, Eve and background noise. Therefore, Fig.~\ref{fig:oursystem} visualises a component of Eve's contribution. In particular, Fig.~\ref{fig:oursystem} visualises the beamsplitter model for the click probabilities for the idler detectors $\mathrm{Pr}_\mathrm{I:\mathcal{X}_+}$ and the click probabilities for Eve's detectors $\mathrm{Pr}_{\mathrm{S:\mathcal{Y}_+\vert I:\mathcal{X}_+}}(\theta_\mathrm{T})$. The system structure shown in Fig.~\ref{fig:oursystem} is equivalent to the scenario where the signal mode light is not intercepted by Eve if her detectors are replaced by Alice's signal H/V detectors and the correct basis decision is always made. In that situation Fig.~\ref{fig:oursystem} instead visualises the beamsplitter model for the click probabilities for the idler detectors $\mathrm{Pr}_\mathrm{I:\mathcal{X}_+}$ and the click probabilities for Alice's signal detectors when her untampered light is incident $\mathrm{Pr}_{\mathrm{S:\mathcal{Z}_+\vert I:\mathcal{X}_+}}$. Moreover, Fig.~\ref{fig:oursystem} shows that background noise is modelled by mixing Alice's light state and the background noise state at a beamsplitter and where the signal loss is encoded by the transmission parameter of the beamsplitters.
Figure~\ref{fig:evesystem} is a beamsplitter diagram which shows Alice's signal detectors when Eve's light is incident. Here, Alice's basis choice is predetermined from her idler measurement. Figure~\ref{fig:evesystem} is similar to Fig.~\ref{fig:oursystem} in that it visualises a sub-system of Eve's contribution to the real/false channel click probabilities. In particular, Fig.~\ref{fig:evesystem} visualises the beamsplitter model which we use to calculate the click probabilities for Alice's signal detectors which has Eve's light incident $\mathrm{Pr}_\mathrm{S:\mathcal{Z}_+\vert E:\mathcal{Y}}(\theta_\mathrm{T})$. Appendix~\ref{appendix:parameters} details and groups the parameters relevant to the system, as visualised in both Fig.~\ref{fig:oursystem} and Fig.~\ref{fig:evesystem}. 

The attenuation of Alice's untampered light from the interaction with a target object is encoded by the signal attenuation factor $\xi$, with $\xi=1$ signifying no attenuation. This is because the (possible) target object is situated in the direct pathway between Alice's light source and Alice's signal detectors: a route untampered by Eve. Whereas, the signal attenuation factor $\xi_\mathfrak{E}$ is set by Eve and does not involve the target object. This is because, for our system, the target object is not situated in-between Alice light source and Eve or in-between Eve and Alice's signal detectors. The geometry of this object detection and range finding system is depicted in Fig.~\ref{fig:rangefinding_sys}. Moreover, as shown in Appendix~\ref{appendix:parameters}, there could be a different signal attenuation factor $\xi_\mathfrak{E}$ for the light Eve resends towards the real and/or false channel from the signal attenuation factor of Alice's untampered light $\xi$.
For simplicity, we assume that the idler and Eve's detectors have minimal background noise incident. Therefore, we can disregard double coincidence clicks at each idler or each of Eve's detector. This assumption is justified if the idler detector system is shielded from the environment and if Eve's detector system is not subject to a jamming-attack which Alice's signal detectors may suffer from. Moreover, the inclusion of substantial background noise at Eve's detectors does not qualitatively affect how this protocol functions as Eve's policy of resending any light she measures means it will simply increase the background noise Alice experiences. We refer to a double (coincidence) click as an event in which both detectors corresponding to the eigenstates of the basis register a click. However, unlike for the idler and Eve's detectors, we cannot disregard the influence of double clicks for Alice's signal detectors: which can have substantial background noise incident. The double clicks include both correct and wrong coincidence clicks. The signal detector double click probability conditioned by an idler click (for polarisation $\mathcal{X}$) is $\mathrm{Pr}_{\mathrm{S:\mathcal{Z}_1,\mathcal{Z}_2}\vert \mathrm{I}:\mathcal{X}}$ and the signal detector double click probability resulting from the light sent by Eve (for polarisation $\mathcal{Y}$) is $\mathrm{Pr}_{\mathrm{S:\mathcal{Z}_1,\mathcal{Z}_2}\vert\mathrm{ E}:\mathcal{Y}}(\theta_\mathrm{T})$. Where $\mathcal{Z}_{1},\mathcal{Z}_{2}\in\{\mathcal{Z}_{+}\} \;\mathrm{ or }\;\{\mathcal{Z}_{\times}\}$.

We now denote the Eve and Alice contribution click probabilities as follows. The Alice/Eve ($\mathfrak{A}/\mathfrak{E}$) correct/wrong/double coincidence click probability is $\mathrm{Pr}^{\mathfrak{A}/\mathfrak{E}}_{\mathrm{c/w/wc}}(\theta)$, where $\theta$ is the relative basis angle as detailed in Sect.~\ref{sect:prelim}. The background noise correct/wrong/double coincidence click probability is $\mathrm{Pr}^{\mathrm{B}}_\mathrm{c/w/wc}$. These are fully defined in Appendix~\ref{appendix:components}. The background noise click probabilities for the real and false channels in principle could be different; however, we shall assume that they are equal for each channel.

However, it is the real or false channel that Alice measures, as the exact contribution of Alice and Eve are in practice unknown. Hence, we define the real/false channel $\mathfrak{R}/\mathfrak{F}$ correct/wrong/double coincidence click probability as $\mathrm{Pr}^{\mathfrak{R}/\mathfrak{F}}_{\mathrm{c/w/wc}}(\theta)$. We are able to state the detection channel click probabilities in terms of the Alice and Eve contributions (and vice-versa). It is these click probabilities that we use to analyse the system henceforth. The correct/wrong/double click probabilities for the real and false channels, in terms of Alice, Eve and noise contributions are respectively \begin{equation}\begin{split}
    \mathrm{Pr}^\mathfrak{R}_\mathrm{c/w/wc}(\theta)&=(1-p)\mathrm{Pr}^\mathfrak{A}_\mathrm{c/w/wc}+p_\mathfrak{R}\mathrm{Pr}^\mathfrak{E}_\mathrm{c/w/wc}(\theta)+ \\ &+p_\mathfrak{F}\mathrm{Pr}^\mathrm{B}_\mathrm{c/w/wc}, \label{eq:real_defn}\end{split}
\end{equation}
\begin{equation}
    \mathrm{Pr}^\mathfrak{F}_\mathrm{c/w/wc}(\theta)=p_\mathfrak{F}\mathrm{Pr}^\mathfrak{E}_\mathrm{c/w/wc}(\theta)+(1-p_\mathfrak{F})\mathrm{Pr}^\mathrm{B}_\mathrm{c/w/wc}.
\label{eq:false_defn}\end{equation} From the above equations, it is apparent that we have assumed that Eve's set of system parameters do not differ whether she sends light towards the real or false channel. It would be straightforward to adjust our definitions of the real and false channel click probabilities if the background noise and Eve system parameters differed between channels; however, this would complicate the following error analysis. Appendix~\ref{appendix:parameters} details that each polarisation mode has its own set of parameters, such as background noise and detector quantum efficiency. Hence, unless stated otherwise, for the remainder of this paper we assume that all of Alice's polarisation modes share the same set of parameters. Similarly, all of Eve's polarisation modes share a different set of parameters. Due to this assumption, the relative basis angle $\theta$ is redundant as there is no difference in the detection statistics with varying the relative basis angle.
\subsection{System security}
\label{sect:prelim}
In this section we describe how security is provided in our spoofing-resilient QI-based object detection and range finding protocol. As stated in the introduction, the security of our detection protocol is provided by the information advantage over Eve. The basis choice information advantage that Alice has is derived from the property of a basis set being mutually unbiased. For example, if Alice encodes a single-photon in an eigenstate in one basis of the set of mutually unbiased bases (MUB), then the measurement result in another basis within the set of MUB is randomly (and equally) distributed between the eigenstates of the measurement basis. Therefore, the information disadvantage that Eve has is due to her lack of knowledge of the selected basis. Eve's intercept-resend strategy can (on average) introduce errors to Alice's correlation measurements due to her wrong basis selection for the intercepted light. The consequent induced error in Alice's correlation measurements is the signature of the presence of Eve. It is in Eve's interest to minimise these wrong correlations. It is also implicit that Alice has to randomly change her basis, otherwise her security is based upon obscurity: which is highly vulnerable to becoming compromised if Eve was to discover that constant basis choice. We also assume that Eve is not technologically limited, that is they are able to produce a single-photon on demand in the degree of freedom of choice and that her measure and resend system can be effectively immediate.

Weaknesses in the security of our protocol can arise from the presence of multi-photons in Alice's source light. Eve can exploit this weakness with attacks such as described by Curty et. al \cite{Curty04}. However, as it is well-known that the benefit of QI-based protocols is more pronounced in the weak source mean photon number regime $\bar{n}\ll 1$. Hence, we assume that Alice's light is weak enough such that multi-photon contributions are negligible. Consequently, Alice's signal light only contains vacuum and single-photon contributions and is immune from multi-photon-based attacks from Eve.

We will focus on the polarisation degree of freedom to realise the photonic encoding. In this paper, for Alice's MUB she uses the rectilinear basis and the diagonal basis. In terms of the bosonic creation operator $\hat{a}^\dag$ the elements of Alice's rectilinear basis consists of the horizontal $\hat{a}^\dag_{\mathrm{H}}$ and vertical $\hat{a}^\dag_{\mathrm{V}}$ polarisations and the elements of Alice's diagonal basis consists of the diagonal $\hat{a}^\dag_{\mathrm{D}}=\frac{1}{\sqrt{2}}\left(\hat{a}^\dag_{\mathrm{H}}+\hat{a}^\dag_{\mathrm{V}}\right)$ and anti-diagonal $\hat{a}^\dag_{\mathrm{A}}=\frac{1}{\sqrt{2}}\left(-\hat{a}^\dag_{\mathrm{H}}+\hat{a}^\dag_{\mathrm{V}}\right)$ polarisations. We focus on polarisation as a degree of freedom as it is easy for Alice to produce a light source which has equal weighting between the elements of each basis for polarisation. However, in practice, polarisation could be unsuitable for object detection and range finding as the reflection coefficient of the target object for different polarisations may be different, depolarisation during transmission through an intermediate medium could occur or more significantly  polarisation may not be preserved on reflection, even for normal incidence. However, if the target is a cooperative one then this may not be such an issue. Otherwise we could move to alternative degrees of freedom such as frequency or time-bin encoding.

\begin{center}\begin{figure}
\includegraphics{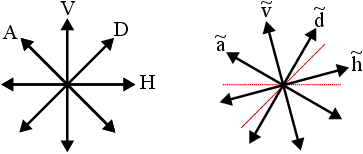}
\caption{Visual comparison of Alice's MUB to the MUB of Eve, for a relative basis angle $0<\theta<\frac{\pi}{4}$.\label{fig:MUBs}}
\end{figure}
\end{center}

Eve has her own MUB, the rectilinear ($\tilde{\mathrm{h}}$ and $\tilde{\mathrm{v}}$) basis and diagonal ($\tilde{\mathrm{d}}$ and $\tilde{\mathrm{a}}$) basis. However, her bases can differ from Alice's by a relative angle $\theta$. For example, the relation of her MUB to Alice's is \begin{align}
   \nonumber \hat{a}^\dag_{\tilde{\mathrm{h}}}&=\mathrm{cos}(\theta)\hat{a}^\dag_{\mathrm{H}}+\mathrm{sin}(\theta)\hat{a}^\dag_{\mathrm{V}}, \\
    &=\mathrm{cos}\left(\frac{\pi}{4}-\theta\right)\hat{a}^\dag_{\mathrm{D}}-\mathrm{sin}\left(\frac{\pi}{4}-\theta\right)\hat{a}^\dag_{\mathrm{A}},\\
    \nonumber \hat{a}^\dag_{\tilde{\mathrm{v}}}&=-\mathrm{sin}(\theta)\hat{a}^\dag_{\mathrm{H}}+\mathrm{cos}(\theta)\hat{a}^\dag_{\mathrm{V}}, \\
    &=\mathrm{sin}\left(\frac{\pi}{4}-\theta\right)\hat{a}^\dag_{\mathrm{D}}+\mathrm{cos}\left(\frac{\pi}{4}-\theta\right)\hat{a}^\dag_{\mathrm{A}}, \\
    \nonumber \hat{a}^\dag_{\tilde{\mathrm{d}}}&=\mathrm{cos}(\theta)\hat{a}^\dag_{\mathrm{D}}+\mathrm{sin}(\theta)\hat{a}^\dag_{\mathrm{A}}, \\
    &=\mathrm{cos}\left(\frac{\pi}{4}-\theta\right)\hat{a}^\dag_{\mathrm{V}}+\mathrm{sin}\left(\frac{\pi}{4}-\theta\right)\hat{a}^\dag_{\mathrm{H}}, \\
    \nonumber \hat{a}^\dag_{\tilde{\mathrm{a}}}&=-\mathrm{sin}(\theta)\hat{a}^\dag_{\mathrm{D}}+\mathrm{cos}(\theta)\hat{a}^\dag_{\mathrm{A}},\\ 
    &=\mathrm{sin}\left(\frac{\pi}{4}-\theta\right)\hat{a}^\dag_{\mathrm{V}}-\mathrm{cos}\left(\frac{\pi}{4}-\theta\right)\hat{a}^\dag_{\mathrm{H}}. \label{eq:basis_relations}
\end{align} We refer to the argument of the trigonometric functions as the total angle $\theta_\mathrm{T}$. This allows for both a possible $\frac{\pi}{4}$ shift and the relative basis angle $\theta$ to be considered when mapping from one basis to another. For example, if the relative basis angle $\theta=0$, then the total angle $\theta_\mathrm{T}=\frac{\pi}{4}$ is used to relate Eve's basis element $\tilde{\mathrm{a}}$ to Alice's basis elements $\mathrm{H}$ and $\mathrm{V}$. Conversely, the total angle $\theta_\mathrm{T}=0$ is used to relate Eve's basis element $\tilde{\mathrm{a}}$ to Alice's basis elements D and A. The relations of her MUB to Alice's is also visualised in Fig.~\ref{fig:MUBs}.

\subsection{Error analysis}
\label{sect:error_analysis}
The intrusion of Eve can introduce errors on average in Alice's detection statistics. If Eve only partially attacks ($p<1$) then this reduces Alice's ability to recognise Eve's intrusion and hence counter her spoofing attempt. This section quantifies the error that Eve induces. In essence, for successful spoofing, Eve should make Alice believe that the false channel represents the true scene, instead of the real channel. In doing so she must ensure that the number of correct coincidence counts for the false channel is equal to or surpasses that for the real channel. However, in regimes without much double click influence, an increase of correct coincidence counts also increases the signature of her intrusion: wrong coincidence counts. A wise strategy for Eve is to resend some light towards the real channel to artificially raise the wrong coincidences; doing such will complicate Alice's attempts to exclude the false channel as it is no longer the sole channel containing the signature of Eve's intrusion. 

We find a factor $k$ which relates the correct coincidence click probability induced by Eve to the wrong coincidence probability. We also show that double click contributions can modify this relation. Moreover, we assume that the background noise is sufficiently static such that there can be accurate knowledge of it. We define $k$ as \begin{equation}
    k=\frac{\mathrm{Pr}^{\mathfrak{E}}_\mathrm{c}-\mathrm{Pr}^{\mathrm{B}}_\mathrm{c}}{\mathrm{Pr}^{\mathfrak{E}}_\mathrm{w}-\mathrm{Pr}^{\mathrm{B}}_\mathrm{w}}.
\end{equation} For the following analysis, we assume that $k\geq1$ and $\mathrm{Pr}^{\mathfrak{E}}_\mathrm{w}-\mathrm{Pr}^{\mathrm{B}}_\mathrm{w}>0$. As mentioned earlier, in general Alice does not know the Eve contribution click probabilities, but she can do a validity check that $\mathrm{Pr}^{\mathfrak{E}}_\mathrm{w}(\xi)-\mathrm{Pr}^{\mathrm{B}}_\mathrm{w}>0$ is true for any $0<\xi\leq 1$. Instead, if $\mathrm{Pr}^{\mathfrak{E}}_\mathrm{w}(\xi)-\mathrm{Pr}^{\mathrm{B}}_\mathrm{w}\leq 0$ for any $0<\xi\leq 1$, Alice is particularly vulnerable to spoofing as double clicks contributions have caused the signature of Eve's intrusion to be obscured by reducing the wrong coincidence click probability below what is expected from noise only. 

Once the $k$-factor is calculated, the error (pertaining to Eve's noise-reduced contributions only) is \begin{equation}
    e_\mathfrak{E}=\frac{1}{k+1}.\label{eq:eve_error} 
\end{equation}The minimum of Eve's contribution error in previous literature is $e_\mathfrak{E}=0.25$ \cite{Bennett92}. Our analysis demonstrates this result for negligible double click regimes. However, we demonstrate that the influence of double clicks means that it is possible for $e_\mathfrak{E}<0.25$. J. Heo et. al provide a threshold error rate $e_\mathrm{T}$ which instructs the minimal error to recognise spoofing in a partial attack \cite{Heo23,Heo24}. If Eve's interception is complete ($p=1$) then $e_\mathrm{T}=e_\mathfrak{E}$, otherwise for a partial spoofing attack ($p<1$) the threshold error is lower $e_\mathrm{T}<e_\mathfrak{E}$.  We define the (noise-reduced) threshold error as
\begin{equation}
    e_\mathrm{T}=e_\mathfrak{E}\frac{p(\mathrm{Pr}^\mathfrak{E}-\mathrm{Pr}^{\mathrm{B}})}{p(\mathrm{Pr}^\mathfrak{E}-\mathrm{Pr}^{\mathrm{B}})+(1-p)(\mathrm{Pr}^\mathfrak{A}-\mathrm{Pr}^{\mathrm{B}})},
\end{equation} where $\mathrm{Pr}^\mathfrak{E}=\mathrm{Pr}^\mathfrak{E}_\mathrm{w}+\mathrm{Pr}^\mathfrak{E}_\mathrm{c}$, $\mathrm{Pr}^\mathfrak{A}=\mathrm{Pr}^\mathfrak{A}_\mathrm{w}+\mathrm{Pr}^\mathfrak{A}_\mathrm{c}$ and $\mathrm{Pr}^\mathrm{B}=\mathrm{Pr}^\mathrm{B}_\mathrm{w}+\mathrm{Pr}^\mathrm{B}_\mathrm{c}$. As Alice does not know the click probabilities, she instead must estimate the threshold error $e_\mathrm{T}$ from her detection statistics of the channels that she investigates.

For unknown click probabilities and proportions of intrusion, estimation of the threshold error is not straightforward as Alice cannot isolate the contributions from Eve and herself from the total detection statistics of the real and false channels. Due to this, there is an off-set between what Alice can measure and the threshold error; hence, we define the offset threshold error $e_\mathrm{T,off}$ as \begin{equation}
    e_\mathrm{T,off}=e_\mathrm{T}+\mathrm{offset},
\end{equation} where \begin{equation}\mathrm{offset}=\frac{(1-p)(\mathrm{Pr}^{\mathfrak{A}}_\mathrm{w}-\mathrm{Pr}^{\mathrm{B}}_\mathrm{w})}{p(\mathrm{Pr}^\mathfrak{E}-\mathrm{Pr}^{\mathrm{B}})+(1-p)(\mathrm{Pr}^\mathfrak{A}-\mathrm{Pr}^{\mathrm{B}})}.\end{equation} If the idler detector background noise mean photon number $\mathbf{\bar{n}_\mathrm{B,I}}\approx \mathbf{0}$, then this ensures that Alice's noise-reduced wrong coincidence probability is $\mathrm{Pr}^\mathfrak{A}_\mathrm{w}-\mathrm{Pr}^{\mathrm{B}}_\mathrm{w}\lessapprox 0$, where less than zero is due to the influence of double clicks. Alice estimates the off-set threshold error from her detection statistics as
\begin{widetext}
 \begin{equation}
    \hat{e}_\mathrm{T,off}=\frac{(\hat{\mathrm{Pr}}^\mathfrak{F}_\mathrm{w}-\mathrm{Pr}^\mathrm{B}_\mathrm{w})+(\hat{\mathrm{Pr}}^\mathfrak{R}_\mathrm{w}-\mathrm{Pr}^\mathrm{B}_\mathrm{w})}{(\hat{\mathrm{Pr}}^\mathfrak{F}_\mathrm{w}-\mathrm{Pr}^\mathrm{B}_\mathrm{w})+(\hat{\mathrm{Pr}}^\mathfrak{R}_\mathrm{w}-\mathrm{Pr}^\mathrm{B}_\mathrm{w})+(\hat{\mathrm{Pr}}^\mathfrak{F}_\mathrm{c}-\mathrm{Pr}^\mathrm{B}_\mathrm{c})+(\hat{\mathrm{Pr}}^\mathfrak{R}_\mathrm{c}-\mathrm{Pr}^\mathrm{B}_\mathrm{c})},\label{eq:measured_e_T}
\end{equation}
\end{widetext} where $\hat{\mathrm{Pr}}^{\mathfrak{R/F}}_\mathrm{w/c}$ is the measured real/false channel wrong/correct coincidence click probability. Estimation of $k$ is straight-forward if there is a channel which has none of Alice's light, i.e. the false channel. However, Alice does not know a priori which (or if a) channel only contains light from Eve, therefore for both channels Alice estimates the respective $k$-factors, if possible. When $p_\mathfrak{F}>0$ Alice estimates $k$ as \begin{equation}
    \hat{k}_\mathfrak{F}=\frac{\hat{\mathrm{Pr}}^{\mathfrak{F}}_\mathrm{c}-\mathrm{Pr}^{\mathrm{B}}_\mathrm{c}}{\hat{\mathrm{Pr}}^{\mathfrak{F}}_\mathrm{w}-\mathrm{Pr}^{\mathrm{B}}_\mathrm{w}}.
\end{equation} Granted accurate estimation, from our definition of the false channel it is clear that $\hat{k}_\mathfrak{F}=k$, hence its suitability. If $p_\mathfrak{R}>1$ Alice estimates the real channel $k$-factor as \begin{equation}
    \hat{k}_\mathfrak{R}=\frac{\hat{\mathrm{Pr}}^\mathfrak{R}_\mathrm{c}-\mathrm{Pr}^{\mathrm{B}}_\mathrm{c}}{\hat{\mathrm{Pr}}^\mathfrak{R}_\mathrm{w}-\mathrm{Pr}^{\mathrm{B}}_\mathrm{w}},
\end{equation} where we assume $\hat{k}_\mathfrak{R}\geq 1$ and $\hat{\mathrm{Pr}}^\mathfrak{R}_\mathrm{w}-\mathrm{Pr}^{\mathrm{B}}_\mathrm{w}>0$. Later, we shall show how Alice can decide which channel (if any) contains light only from Eve and which channel (if any) contains her own light, which is possibly mixed with Eve's light.

However, if $p_\mathfrak{F}=0$ accurate estimation of $k$ must follow from an alternative method.  Alice can estimate $k$ by calculating \begin{equation}k(0<\xi_\mathfrak{E}\leq 1)=\frac{\mathrm{Pr}^{\mathfrak{E}}_\mathrm{c}(0<\xi_\mathfrak{E}\leq 1)-\mathrm{Pr}^{\mathrm{B}}_\mathrm{c}}{\mathrm{Pr}^{\mathfrak{E}}_\mathrm{w}(0<\xi_\mathfrak{E}\leq 1)-\mathrm{Pr}^{\mathrm{B}}_\mathrm{w}},\end{equation} for range of values found by $0<\xi\leq 1$. If there is a small relative error between $k(\xi_\mathfrak{E}\gtrapprox 0)$ and $k(\xi_\mathfrak{E}=1)$, then Alice is able to estimate Eve's contribution error $e_\mathfrak{E}$. Although this method depends upon Alice knowing the system parameters (except for $\xi_\mathfrak{E}$) for Eve's detection system.

\subsection{Recognition of Eve}
\label{sect:recognition_eve}
In general, Alice does not know the signal detector click probabilities due to a lack of knowledge of the target object properties and/or the parameters used by Eve, therefore Alice depends upon what she measures for recognition of Eve's intrusion. By convention, we continue to refer to the channels as real and false, even if Alice does not know which is which a priori, or even if the real channel contains any of Alice's untampered light. This is because we define them from Eq.~\ref{eq:real_defn} and Eq.~\ref{eq:false_defn}, where the real channel has the possibility of Alice's untampered light whereas the false channel does not. Prior to any decision-making about the presence of Eve and/or Alice's light, Alice checks if a channel $\mathfrak{G}$ has both noise-reduced wrong and correct coincidence click expectation values $\hat{\mathrm{Pr}}^\mathfrak{G}_\mathrm{w/c}-\mathrm{Pr}^{\mathrm{B}}_\mathrm{w/c}=0$. If this is true, then this absence of (additional) light infers that there is no target present there and/or Eve's light is entirely sent towards another channel. 

Estimation of error values can help Alice recognise the presence of Eve. Alice initiates her attempt to recognise Eve's intrusion by checking if Eq.~\ref{eq:measured_e_T} (the measured offset threshold error) is greater than zero. If this is satisfied Alice has recognised the presence of Eve, but she has not said which channel(s) contains light from Eve. If the offset threshold error is $e_\mathrm{T,off}\leq 0$ it is not as easy to recognise Eve's intrusion; Alice must instead look for discrepancies in the measured statistics. This approach is weaker for successful decision-making as it relies upon two layers of estimation. The discrepancy check method goes as follows. First, Alice checks if any channel has a positive noise-reduced wrong coincidence click expectation value: this infers the intrusion of Eve in that channel. If it is ruled out that any channel has a positive noise-reduced wrong coincidence click expectation value Alice uses the discrepancy method to check a channel which has a negative or approximately zero noise-reduced wrong coincidence click expectation value. For a channel $\mathfrak{G}$ which has a negative or approximately zero noise-reduced wrong coincidence click expectation value $\hat{\mathrm{Pr}}^\mathfrak{G}_\mathrm{w}-\mathrm{Pr}^{\mathrm{B}}_\mathrm{w}\lessapprox 0$ we assume that the correct coincidence click probability only contains Alice and noise contributions \begin{equation}\hat{\mathrm{Pr}}^\mathfrak{G}_\mathrm{c}=(1-p^{\mathfrak{G}})\mathrm{Pr}^\mathfrak{A}_\mathrm{c}(\xi^\mathfrak{G})+p^\mathfrak{G}\mathrm{Pr}^{\mathrm{B}}_\mathrm{c},\label{eq:equate_real_with_alice}\end{equation} Alice now calculates the set of solutions with probability of channel $\mathfrak{G}$ intrusion $p^\mathfrak{G}$ and signal attenuation factor $\xi^\mathfrak{G}$ which satisfies Eq.~\ref{eq:equate_real_with_alice}. If there was no intrusion from Eve then the set of solutions $(p^\mathfrak{G},\xi^\mathfrak{G})$ will also satisfy $(1-p^\mathfrak{G})\mathrm{Pr}^\mathfrak{A}_\mathrm{w}(\xi^\mathfrak{G})+p^\mathfrak{G}\mathrm{Pr}^{\mathrm{B}}_\mathrm{w}=\hat{\mathrm{Pr}}^\mathfrak{G}_\mathrm{w}$ and if there was intrusion from Eve then $(1-p^\mathfrak{G})\mathrm{Pr}^\mathfrak{A}_\mathrm{w}(\xi^\mathfrak{G})+p^\mathfrak{G}\mathrm{Pr}^{\mathrm{B}}_\mathrm{w}\neq\hat{\mathrm{Pr}}^\mathfrak{G}_\mathrm{w}$. This discrepancy check method requires that $\mathrm{Pr}^\mathfrak{E}_\mathrm{w}-\mathrm{Pr}^\mathrm{B}_\mathrm{w}>0$.

Once Alice has confirmed the intrusion or lack of intrusion from Eve on a channel she proceeds to determine whether or not there is a channel which contains her light (or not). It is this decision-making process which gives Alice a more potent form of spoofing resilience, the knowledge of which channel her trustworthy signal dwells in.  

We begin with situations where the intrusion is complete $p=1$, therefore $\hat{e}_\mathrm{off,T}=e_\mathfrak{E}$ and Alice can conclude no channel contains her light. If $0<p_\mathfrak{{R}}<1$ and $0<p_\mathfrak{F}<1$, then the real and false channel k-factors are equal $\hat{k}_\mathfrak{R}=\hat{k}_\mathfrak{F}$ and both only include Eve's light. Alice can conclude that no channel contains her own light. If $p=p_\mathfrak{F}=1$, then the real channel contains no additional light and the false channel only contains Eve's light, which leads Alice to the conclusion that no channel contains her light as $\hat{e}_\mathrm{off,T}=e_\mathfrak{E}$. The same situation follows if $p=p_\mathfrak{R}=1$, but instead the real channel contains Eve's light only. If an object is absent and there is complete intrusion $p=1$ or incomplete intrusion $p<1$, the same conclusions apply as the previous three examples, correspondingly. In other words, Alice is unable to distinguish between complete intrusion with an object present and complete/incomplete intrusion when an object is absent. Alice can recognise the presence of Eve, but is unable to access any trustworthy signal. The remainder of our discussion about what conclusions Alice can make for different scenarios is only for when a target object is present.

Alice has the chance to decide which or if a channel contains her own light when the intrusion by Eve is incomplete $p<1$ and hence the offset threshold error is $\hat{e}_\mathrm{off,T}<e_\mathfrak{E}$. When $p_\mathfrak{F}\neq p,0$ and $p_\mathfrak{R}\neq p,0$ then both the real channel and false channel contains light from Eve. If $\hat{\mathrm{Pr}}^\mathfrak{R/F}_\mathrm{w}-\mathrm{Pr}^{\mathrm{B}}_\mathrm{w}>0$ is true for both real and false channels then Alice decides that her light dwells in the channel with the larger $k$-factor, as $k_\mathfrak{R}>k_\mathfrak{F}$, as proven in Appendix~\ref{appendix:proof}. Otherwise, if one channel has $\hat{\mathrm{Pr}}^\mathfrak{R}_\mathrm{w}-\mathrm{Pr}^{\mathrm{B}}_\mathrm{w}<0$, then Alice knows automatically that this channel contains her own light, due to the $\hat{\mathrm{Pr}}^\mathfrak{F}_\mathrm{w}-\mathrm{Pr}^{\mathrm{B}}_\mathrm{w}>0$ condition (implicitly) made earlier. If the intrusion is focused entirely on the false channel $p_\mathfrak{F}=p$ it is easy to conclude that the real channel has Alice's light as it is undisturbed, i.e. there is no discrepancy when Alice tests that channel for its wrong coincidence click probability. If the intrusion is focused entirely on the real channel $p_\mathfrak{R}=p$ Alice can see that the false channel has no additional light and she can exclude it. If there is a negative noise-reduced wrong coincidence click expectation value Alice knows that the real channel contains her own light. If there is a positive noise-reduced wrong coincidence click expectation value and Alice knows Eve's contribution error $e_\mathfrak{E}$, Alice concludes that the real channel contains her own light as $\hat{e}_\mathrm{off,T}<e_\mathfrak{E}$.

The above analysis depends upon the requirement that $\mathrm{Pr}^{\mathfrak{E}}_\mathrm{w}(0<\xi_\mathfrak{E}\leq 1)-\mathrm{Pr}^{\mathrm{B}}_\mathrm{w}>0$, otherwise Alice is vulnerable to spoofing. We now describe a scenario where this requirement is not satisfied (due to excessive background noise). The background noise in this scenario is not strong enough to constitute jamming as in theory Alice could detect the object if she knew which channel was real. Less background noise is required to ensure successful spoofing than successful jamming as the wrong coincidence count is smaller (and hence more vulnerable to being siphoned off by double clicks) than the correct coincidence clicks which enable object detection (the complete siphoning of correct coincidence clicks to double clicks would mean Eve has jammed us). For the parameter regimes such that $\mathrm{Pr}^{\mathfrak{F}}_\mathrm{w}-\mathrm{Pr}^{\mathrm{B}}_\mathrm{w}<\mathrm{Pr}^{\mathfrak{R}}_\mathrm{w}-\mathrm{Pr}^{\mathrm{B}}_\mathrm{w}$ and $\mathrm{Pr}^{\mathfrak{F}}_\mathrm{c}-\mathrm{Pr}^{\mathrm{B}}_\mathrm{c}>\mathrm{Pr}^{\mathfrak{R}}_\mathrm{c}-\mathrm{Pr}^{\mathrm{B}}_\mathrm{c}$ this would mean that Alice would be more inclined to believe that the false channel contains her own light rather than the real channel. In other words, Alice has been spoofed.

\subsection{Removal of false information}
\label{sect:removal}
Alice can remove false information if she knows which channel is false and that there is corresponding positive $k$-factor. For example, the noise-reduced trustworthy false channel $\mathfrak{F}^{'}$ correct coincidence click expectation value is \begin{equation}
    \mathrm{Pr}^{\mathfrak{F}^{'}\vert k_\mathfrak{F}}=(\mathrm{Pr}^\mathfrak{F}_\mathrm{c}-\mathrm{Pr}^\mathrm{B}_\mathrm{c})-k_\mathfrak{F}(\mathrm{Pr}^\mathfrak{F}_\mathrm{w}-\mathrm{Pr}^\mathrm{B}_\mathrm{w}),\label{eq:false_k_false}
\end{equation} where from the above definitions it is clear that $\mathrm{Pr}^{\mathfrak{F}^{'}\vert k_\mathfrak{F}}=0$, i.e. all of Eve's contribution has been removed and the background noise accounted for. If the real channel wrong noise-reduced coincidence click expectation value is $\mathrm{Pr}^\mathfrak{R}_\mathrm{w}-\mathrm{Pr}^{\mathrm{B}}_\mathrm{w}>0$ (because $p_\mathfrak{R}>0$) Alice can attempt to remove untrustworthy information from this channel too. The noise-reduced trustworthy real channel $\mathfrak{R}^{'}$ correct coincidence click expectation value is 
\begin{align}
    \mathrm{Pr}^{\mathfrak{R}^{'}\vert k_\mathfrak{F}}&=(\mathrm{Pr}^\mathfrak{R}_\mathrm{c}-\mathrm{Pr}^{\mathrm{B}}_\mathrm{c})-k_\mathfrak{F}(\mathrm{Pr}^\mathfrak{R}_\mathrm{w}-\mathrm{Pr}^{\mathrm{B}}_\mathrm{w}),\nonumber \\
    &=(1-p)(\mathrm{Pr}^\mathfrak{A}_\mathrm{c}-\mathrm{Pr}^{\mathrm{B}}_\mathrm{c})- e(\mathrm{Pr}^{\mathfrak{R}^{'}\vert k_\mathfrak{F}}),\label{eq:false_k_real}
\end{align} where there is an erroneous term
\begin{equation}
    e(\mathrm{Pr}^{\mathfrak{R}^{'}\vert k_\mathfrak{F}})=k_\mathfrak{F}(1-p)(\mathrm{Pr}^\mathfrak{A}_\mathrm{w}-\mathrm{Pr}^{\mathrm{B}}_\mathrm{w}).
\end{equation}

Information removal using the false channel $k$-factor $k_\mathfrak{F}$ results in Eq.~\ref{eq:false_k_false} and Eq.~\ref{eq:false_k_real} for the false and real channel, respectively. This results in the correct coincidence click expectation value of the (noise-reduced) false channel correctly being zero, it also results in the removal of Eve's contribution from the (noise-reduced) real channel correct coincidence click expectation value, albeit perturbed by $e(\mathrm{Pr}^{\mathfrak{R}^{'}\vert k_\mathfrak{F}})$. Here, we have assumed that there is a target object present and $p<1$. 
\subsection{Performance analysis}
\label{sect:perform}
For a set of parameters described in Appendix~\ref{appendix:parameters} we show in Fig.~\ref{fig:relativebasischoice} that there is an optimal choice of relative basis angle for Eve, whereby the optimal choice is the one which which minimises the error rate of Eve's contributions $e_{\mathfrak{E}}(\theta)$. For this example, the optimal relative basis angle is $\theta=0$. Eve could enact this optimisation granted she knows the entire set of parameters relevant to this system and its geometry. However, in general she does not have this information and shall randomly choose a relative basis angle.
\begin{center}\begin{figure}
\includegraphics{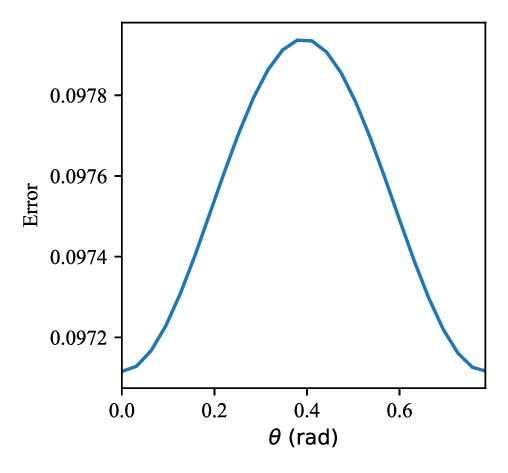}
\caption{The error rate of Eve's (isolated) contribution to Alice's detection statistics as a function of relative basis angle $\theta$. For the parameter regime detailed in Appendix~\ref{appendix:parameters} there is a minimal error for a particular relative basis angle $\theta$. The quantum efficiencies of Eve's and Alice's signal detectors are all different, which causes the variation of the error rate shown. \label{fig:relativebasischoice}}
\end{figure}
\end{center}

\begin{center}
\begin{figure}
\includegraphics{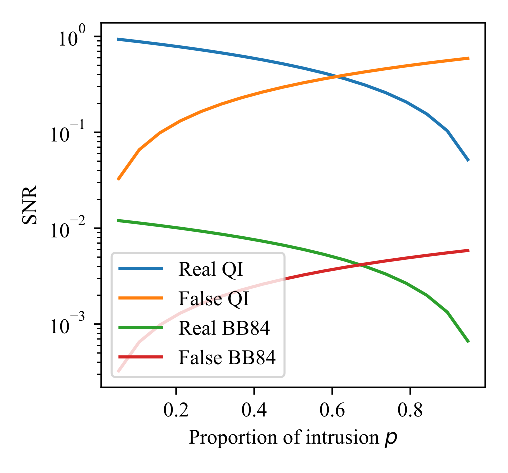}
\caption{SNR for the real and false channels as a function of probability of interception $p$. The parameter regime is detailed in Appendix~\ref{appendix:parameters}. \label{fig:SNRfig}}
\end{figure}
\end{center}

\begin{center}
\begin{figure}
\includegraphics{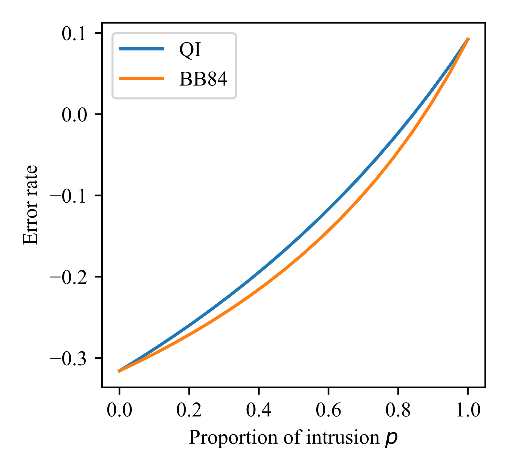}
\caption{Noise-reduced wrong coincidence click expectation value distributions for the real and false channels, for the parameter regime detailed in Appendix~\ref{appendix:parameters}. \label{fig:error_threshold}}
\end{figure}
\end{center}

We now compare Alice's QI light source with a weak-coherent state scenario. The weak-coherent state scenario uses a BB84-inspired prepare-and-measure scheme as described by Malik et. al. \cite{Malik12}. Appendix~\ref{appendix:weak_coherent_state_prob} details the click probabilities used for the analysis with a weak-coherent state as Alice's source light. We define signal to noise (SNR) for the real/false channel as \begin{equation}
    \mathrm{SNR}^{\mathfrak{R/F}}=\frac{\mathrm{Pr}^{\mathfrak{R/F}}_{\mathrm{c}}-\mathrm{Pr}^{\mathrm{B}}_{\mathrm{c}}+\mathrm{Pr}^{\mathfrak{R/F}}_{\mathrm{w}}-\mathrm{Pr}^{\mathrm{B}}_{\mathrm{w}}}{\mathrm{Pr}^{\mathrm{B}}_{\mathrm{c}}+\mathrm{Pr}^{\mathrm{B}}_{\mathrm{w}}}.
\end{equation}
Figure~\ref{fig:SNRfig} shows the QI and BB84 SNR for the real and false channels as a function of probability of interception $p$, where $p=p_\mathfrak{F}$. When $p\approx 0.684$ the QI real and false channels have the same probability of correct coincidence click, this choice of $p$ is wise for Eve as she does not unnecessarily induce extra wrong coincidence clicks. It is also apparent that the BB84-inspired system is inferior due to the low SNR compared to the QI system. Figure~\ref{fig:error_threshold} shows the offset threshold error $e_\mathrm{off,T}$ for QI and BB84. The QI system is marginally better at the recognition of Eve's intrusion via her induced error. However, the noise-reduced technique is less realistic for BB84, as due to its high level of noise it would take far longer to sample an accurate estimate of the noise than it would for QI. Due to the advantage of the QI system over the BB84 one, we continue to focus on the QI one for the remainder of this paper.

\subsection{Spoofing error thresholding}
\label{sect:thresholding}
The above analysis depends upon accurate estimation of the underlying click probabilities from the detection statistics. However, Alice may be in a situation where such an accurate estimation is unavailable due to constrained integration times. In this section we describe the underlying distributions that Alice samples from. At the limit of many samples, our earlier approach for making conclusions about Eve's intrusion and for which channel Alice's signal data lies applies. Whereas from just one sample we show that such conclusions could be invalid due to the variances of the underlying distributions. In particular, the discrepancy check method is infeasible as it requires two accurate estimations.

For non-zero coincidence counts in both channels and when the requirement $\mathrm{Pr}^{\mathfrak{E}}_\mathrm{w}-\mathrm{Pr}^{\mathrm{B}}_\mathrm{w}>0$ is satisfied, an example case of spoofing-vulnerability from limited sampling is that when the estimated false channel noise-reduced wrong coincidence expectancy is negative while the real channel is positive, Alice would be spoofed. From analysis of these distributions Alice can set thresholds on how much the possibility of spoofing is tolerable. If certain parameter regimes are within these spoofing thresholds for any value of the unknown signal attenuation factor, Alice can then ascertain that this parameter regime is spoofing-resilient even from a limited sample size of just one. 

\begin{center}
\begin{figure}
\includegraphics{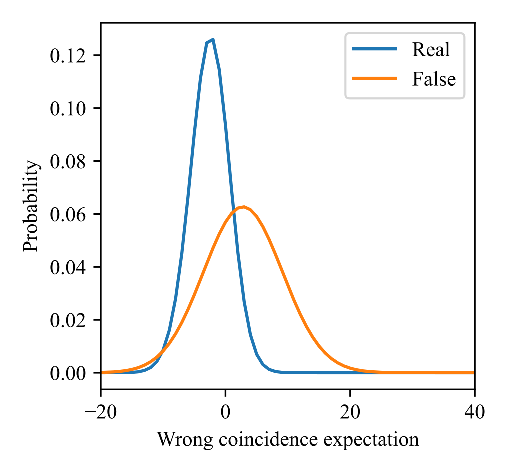}
\caption{Noise-reduced wrong coincidence click expectation value distributions for the real and false channels, for the parameter regime detailed in Appendix~\ref{appendix:parameters}. The noise-reduced wrong coincidence click expectation value is on the x-axis and the probability according to the expectation value is on the y-axis. \label{fig:expectation_dists}}
\end{figure}
\end{center}

The object present distribution for the real and false channel is Poissonian, as is the background noise. The difference of these Poisson distributions is known as the Skellam distribution \cite{Skellam46,karlis03}. There is a covariance for the two distributions (as the object present number of clicks is dependent upon the random variable for the number of background noise clicks). This covariance $C$ is not trivial to find and we shall use our Monte-Carlo (M.C.) simulation for its calculation: the details of our simulation are in Appendix~\ref{appendix:monte_carlo}. Hence, the noise-reduced real channel wrong coincidence click expectation value distribution is defined as \begin{equation}
    P(x,\mu_1,\mu_2)=e^{-\mu_1-\mu_2}\left(\frac{\mu_1}{\mu_2}\right)^{\frac{x}{2}}\mathrm{I}_{\vert x\vert}(2\sqrt{\mu_1\mu_2}),
\end{equation} where $x$ is the expected number of wrong coincidence clicks, $\mu_1=N\;\mathrm{Pr}^{\mathfrak{R}}_\mathrm{w}-C$ and $\mu_2=N\;\mathrm{Pr}^{\mathrm{B}}_\mathrm{w}-C$, with $N$ as the number of shots. Also $\mathrm{I}_{\vert x\vert}$ is the modified Bessel function of the first kind. Figure~\ref{fig:expectation_dists} shows the real and false channel noise-reduced wrong coincidence click expectation value distributions when Eve only intrudes upon the false channel. Even when have made the assumption that $\mathrm{Pr}^{\mathfrak{A}}_\mathrm{w}-\mathrm{Pr}^{\mathrm{B}}_\mathrm{w}<0$ we can see that there is the possibility of positive values for the real channel noise-reduced wrong coincidence click expectation value, which is at odds with the probability analysis section prior.
We can calculate the probability that some conclusions Alice makes about Eve's intrusion and the whereabouts of her own light are erroneous when sampling from a distribution rather than from the accurate expectation values as described in Sect.~\ref{sect:recognition_eve}. An example conclusion Alice can threshold for is the probability that the noise-reduced false channel wrong coincidence expectation value is negative when the noise-reduced real channel wrong coincidence expectation value is positive $\mathrm{Pr}\left(\mathrm{Pr}^{\mathfrak{F}}_\mathrm{w}-\mathrm{Pr}^{\mathrm{B}}_\mathrm{w}<0\; \vert \; \mathrm{Pr}^{\mathfrak{R}}_\mathrm{w}-\mathrm{Pr}^{\mathrm{B}}_\mathrm{w}>0 \right)$. Another example of an erroneous conclusion is when both $\mathrm{Pr}^{\mathfrak{F/R}}_\mathrm{w}-\mathrm{Pr}^{\mathrm{B}}_\mathrm{w}>0$ Alice measures that $k_\mathfrak{F}>k_\mathfrak{R}$. The probability of this is stated as $\mathrm{Pr}(k_\mathfrak{F}>k_\mathfrak{R}>0)$. Therefore, in both of these scenarios Alice erroneously concludes that the false channel contains her own light instead of the real channel. As explained earlier the discrepancy check is not feasible in this limited sampling situation, therefore Alice is unable to make decisions about the presence of Eve or the whereabouts of her own light when she measures the off-set threshold error as $\hat{e}_\mathrm{off,T}\leq 0$.

When Alice measures that $\hat{e}_\mathrm{T,off}\leq 0$ she does not recognise any intrusion by Eve. Erroneous conclusions can be made here with two possibilities. There is the probability that $\mathrm{Pr}\left(\mathrm{Pr}^{\mathfrak{R}}_\mathrm{w}-\mathrm{Pr}^{\mathrm{B}}_\mathrm{w}>0\; \mathrm{and} \; \mathrm{Pr}^{\mathfrak{F}}_\mathrm{w}-\mathrm{Pr}^{\mathrm{B}}_\mathrm{w}<0  \vert \hat{e}_\mathrm{T,off}\leq 0 \right)$. There is also the probability that $\mathrm{Pr}\left(\mathrm{Pr}^{\mathfrak{F}}_\mathrm{w}-\mathrm{Pr}^{\mathrm{B}}_\mathrm{w}\leq \mathrm{Pr}^{\mathfrak{R}}_\mathrm{w}-\mathrm{Pr}^{\mathrm{B}}_\mathrm{w}<0  \vert \hat{e}_\mathrm{T,off}\leq 0 \right)$.

\section{Range finding case example}
\label{sect:rangefinding}
We simulate a situation where Eve attempts to temporally spoof Alice, i.e. spoof Alice's range finding decision-making. For simplicity, we assume that the real channel only includes Alice's untampered light and the signal detector background noise and the false channel only includes the tampered light and signal detector background noise. In other words we set $p_\mathfrak{R}=0$ and $p_\mathfrak{F}=p$. We also assume that Alice effectively sends out the vacuum state to Eve for when Alice registers an idler no-click event. This is valid if the idler detector background noise $\bar{n}_\mathrm{B,I}\approx 0$ and the idler detector quantum efficiency $\eta_\mathrm{I}\approx 1$.

\begin{center}\begin{figure}
\includegraphics{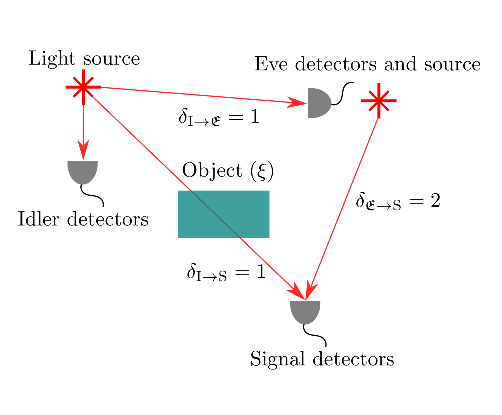}
\caption{Diagram of the pathways (and resulting delays) Alice's light can take when in the presence of Eve undertaking a temporal spoofing attempt. The delay in time is $\delta$ time-bins. The temporal spoofing by Eve delays the light incident upon Alice's signal detector by one time-bin. The influence of a possible target object lies in the direct pathway of Alice's light source and the signal detectors.}\label{fig:rangefinding_sys}
\end{figure}
\end{center}

Figure~\ref{fig:rangefinding_sys} shows how Alice's light is redirected and delayed when Eve attempts temporal spoofing. The assumption that Eve is not technologically limited implies that the quantum efficiency of Eve's detectors are $\eta_\mathfrak{E}\approx 1$. The delay in time is $\delta$, where we have discretised time in terms of time-bins which have the duration of a shot of the system. We can define the duration of a shot either artificially by the coincidence window size of Alice's coincidence counting hardware, or the repetition rate of a pulsed source, for example. Figure~\ref{fig:rangefinding_sys} shows that Eve is attempting to temporally spoof Alice by delaying the return light for the signal detectors by one time-bin ($\delta=1$). Figure~\ref{fig:rangefinding_sys} also shows that the possible target object lies in the direct pathway of Alice's light source and the signal detectors. In our simulation we discard situations where a subsequent idler click initiates a real channel at time-bin $z$ which already has a false channel initiated by a previous idler click. We refer to such events as polluted time-bins: these events are limited granted that $\bar{n}\ll 1$.

\begin{center}\begin{figure}
\includegraphics{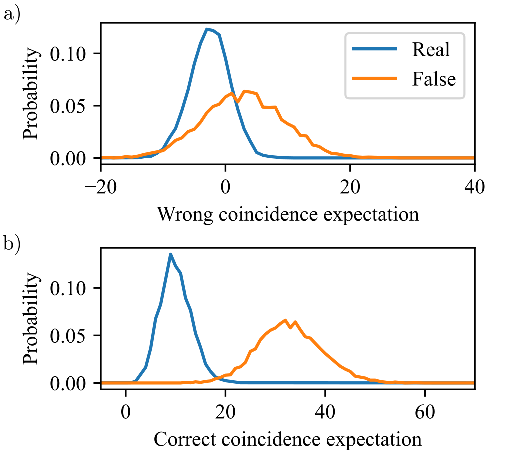}
\caption{a) Shows the noise-reduced wrong coincidence clicks expectation distributions for the real and false channels. The noise-reduced wrong coincidence click expectation value is on the x-axis and probability of that expectation value is on the y-axis. b) Shows the noise-reduced correct coincidence click expectation distributions for real and false. The noise-reduced correct coincidence click expectation value is on the x-axis and probability of that expectation value is on the y-axis. The parameter regime for this figure is detailed in Appendix~\ref{appendix:parameters}.} \label{fig:expectation_hist}
\end{figure}
\end{center}

Figure~\ref{fig:expectation_hist} shows the noise-reduced wrong/correct coincidence click expectations for real and false channels. These distributions have been generated from $5000$ runs of the M.C. simulation described in Appendix~\ref{appendix:monte_carlo}. Inspecting the simulation data we can calculate the probabilities of erroneous conclusions with respect to the number of simulation runs. Alice does not recognise Eve's intrusion with the probability $\mathrm{Pr}(\hat{e}_\mathrm{T,off}\leq 0)=0.5104$. If Alice does not recognise spoofing she could proceed without realising the risk of being spoofed. Our simulation results for the probability of erroneous conclusions, given that spoofing is not recognised, is $\mathrm{Pr}\left(\mathrm{Pr}^{\mathfrak{R}}_\mathrm{w}-\mathrm{Pr}^{\mathrm{B}}_\mathrm{w}>0\; \mathrm{and} \; \mathrm{Pr}^{\mathfrak{F}}_\mathrm{w}-\mathrm{Pr}^{\mathrm{B}}_\mathrm{w}<0  \vert \hat{e}_\mathrm{T,off}\leq 0 \right)=0.0346$ and $\mathrm{Pr}\left(\mathrm{Pr}^{\mathfrak{F}}_\mathrm{w}-\mathrm{Pr}^{\mathrm{B}}_\mathrm{w}\leq \mathrm{Pr}^{\mathfrak{R}}_\mathrm{w}-\mathrm{Pr}^{\mathrm{B}}_\mathrm{w}<0  \vert \hat{e}_\mathrm{T,off}\leq 0 \right)=0.1668$.

Even if Alice recognises Eve's intrusion there is probability that she is spoofed via $\mathrm{Pr}\left(\mathrm{Pr}^{\mathfrak{F}}_\mathrm{w}-\mathrm{Pr}^{\mathrm{B}}_\mathrm{w}<0\; \vert \; \mathrm{Pr}^{\mathfrak{R}}_\mathrm{w}-\mathrm{Pr}^{\mathrm{B}}_\mathrm{w}>0 \right)=0.0222$. Lastly, the probability that Alice is spoofed via $k$-factor analysis is $\mathrm{Pr}(k_\mathfrak{F}>k_\mathfrak{R}>0)=0.07$.
\section{Discussion}
\label{sect:discussion}
In this paper a theory for recognising and accounting for an intercept-resend spoofing attack by an intruder within the context of a simple-detection QI-based LIDAR protocol has been provided. Our model is written in terms of the Fock state formalism for the states of light and detectors involved and we include realistic features such as background noise and signal loss. We also benchmark our system performance with a classical illumination based protocol. This model extends the capabilities of our pre-existing jamming-resilient simple-detection QI-based LIDAR protocol \cite{Murchie24}. 

Security when performing object detection and range finding can be provided when an imager sends their signal light in a degree of freedom which has the property of being an eigenstate of a basis in a set of mutually unbiased bases. There is an information disadvantage that an intruder has as they do not know which basis was chosen by the imager. On average, the intruder will introduce errors in the light they resend (for certain parameter regimes) if they measure the imager's light that they intercept in the wrong basis. As we consider twin beam light - which is highly temporally-correlated - for our object detection and range finding system, the imager inspect the coincidence clicks of their locally sent light (idler) and the light sent towards the possible target object (signal). The imager can recognise that an intruder has attempted a spoofing attack by inspecting their detection statistics for wrong coincidence clicks. Where we define a wrong coincidence click as when the imager measures signal light in an eigenstate which was not correlated with their concurrent (to their chosen delay) idler light. We show that there are sets of system parameters which cause there to be an optimal relative basis angle an intruder can select for a minimised induced error. From the click probabilities that we calculate, we define error rates and signal-to-noise ratios. Analysis with these click probabilities can allow for the recognition of the spoofing attack. This analysis can also allow for finding the whereabouts of the imager's untampered light (if any) and the intruder light (if any). It can also allow for the removal of untrustworthy detector data. Furthermore, we compare the system performance of the QI- and classical illumination- based systems: the results of which reinforce the advantage of QI-based systems in the low signal strength and high background noise parameter regimes that we focus on. 

Certain regimes can allow for a successful spoofing attack without also causing jamming to our object detection protocol. Such spoofing attacks will not be recognised by the conventional error analysis. However, this protocol provides an alternative method for recognising an attack by investigating if there is a discrepancy in the measured click probabilities against our expected click probabilities under the assumption there is no intruder present. In practice, accurate estimation of the system parameters is often impractical and hence the imager is constrained to limited sampling to make conclusions about the object detection and range finding situation. This analysis involves the difference of our object present and the noise-only click-count distributions, which invokes the use of the Skellam distribution. From analysing these distributions we provide thresholds for erroneous conclusions when constrained to limited sampling. This analysis allows for certain parameter regimes to be declared spoofing-resilient, as even from a limited sample the imager is able to ensure that the detector data conclusions are trustworthy. 

As QI-based protocols are most advantageous in constrained signal strength regimes we need not worry about protocol weaknesses that arise from multi-photon contributions: assuming that we are in the weak signal strength limit. However, S. Johnson et. al. describe a potential attack where the imager's idler beam is compromised and an intruder can intercept it \cite{Johnson24}. This attack would alter the arrival time of the imager's idler photon while leaving its polarisation state unmeasured: therefore avoiding inducing any errors. This could lead to temporal spoofing which would be unrecognisable via the error or discrepancy methods. Although, current technology means that the success of this attack is low and would cause a drastic reduction of the imager's coincidence counting.

Further work could involve maximising the number of bases used for photonic encoding. For example, three is the maximum number of MUB for our 2-dimensional system \cite{wooters89}. For the polarisation degree-of-freedom the additional basis that we have not considered comprises of the left/right circular polarisation states. Beyond this, we could extend the dimensionality of our protocol's photonic encoding, which in turn would increase the maximum number of MUB. An increase of the number of MUB could reduce the likelihood that an intruder would successfully spoof the imager, as they would have a relatively larger information disadvantage. Furthermore, this protocol could be modified to have a non-adversarial application in the resilience to self-interference when performing object detection and range finding in complicated scenarios involving multiple targets (with correspondingly multiple signal echos). This could improve range finding as it would be harder for the imager's own return signals from previous time-bins to accidentally overlay with the imager's return signal at a desired delay, as their previous pulse in another basis has reflected from a more distant scatterer, for example. Lastly, in particular, single-photon underwater object detection and range finding could benefit from this resilience to self-interference due to the high level of backscattering by the medium \cite{Maccarone23}.

\section*{Acknowledgements}
The data presented in this work are available at [?]. The authors thank Jonathan D. Pritchard for useful
discussions. The authors would also like to thank the UK Engineering and Physical Sciences Research Council for the partial funding of this work via the UK National Quantum Technology Programme and the QuantIC Imaging Hub (Grant No. EP/T00097X/1).
\appendix
\section{Spoofing click probability}
\label{appendix:spoofing_click_prob_deriv}
\subsection{Parameters}\label{appendix:parameters}
This subsection groups and describes the many parameters in this system. It also states how the parameters are set for each non-schematic figure. As there is a detector for each basis element (polarisation), we can write the idler detector quantum efficiency as a vector $\mathbf{\eta}_\mathrm{I}=\{ \eta_\mathrm{I:H},\eta_\mathrm{I:V},\eta_\mathrm{I:D},\eta_\mathrm{I:A}\}$ and if all four polarisation channels have the same detector quantum efficiency then we can signify this as $\mathbf{\eta}_\mathrm{I}=\mathbf{0.001}$, for example. This style of formatting will follow for the other applicable parameters.

There is also the Eve detector quantum efficiency $\mathbf{\eta}_\mathfrak{E}=\{ \eta_{\mathfrak{E}:\mathrm{\tilde{\mathrm{h}}}},\eta_{\mathfrak{E}:\mathrm{\tilde{\mathrm{v}}}},\eta_{\mathfrak{E}:\mathrm{\tilde{\mathrm{d}}}},\eta_{\mathfrak{E}:\mathrm{\tilde{\mathrm{a}}}}\}$ and Alice's signal detector quantum efficiency $\mathbf{\eta}_\mathrm{S}=\{ \eta_\mathrm{S:H},\eta_\mathrm{S:V},\eta_\mathrm{S:D},\eta_\mathrm{S:A}\}$. The signal attenuation factor for Alice's untampered light is $\xi$ and the signal attenuation factor which Eve sets for the resent light towards Alice is $\xi_\mathfrak{E}$. The signal strength of the BB84 system is the mean photon number $\bar{n}_\alpha$. The mean photon number of the QI system is $\bar{n}$. To keep the comparison between the BB84 and QI system fair, the BB84 mean photon number relates to the QI mean photon number, under the condition of equal idler detector quantum efficiency over all polarisation channels, is $\bar{n}=\frac{\bar{n}_\alpha}{2-\eta_\mathrm{I}[0]\bar{n}_\alpha+2\eta_\mathrm{I}[0]\bar{n}_\alpha}$, where $\eta_\mathrm{I}[0]$ is the first element from the idler detector quantum efficiency vector.

As per our beamsplitter model for incorporating noise and loss, we must rescale the background noise mean photon number by the reflection parameter of the beamsplitter in question. We define the idler detector background noise mean photon number vector as $\bar{n}_\mathrm{B,I}=\{ \frac{\bar{n}_\mathrm{B,I:H}}{1-\eta_\mathrm{I:H}},\frac{\bar{n}_\mathrm{B,I:V}}{1-\eta_\mathrm{I:V}},\frac{\bar{n}_\mathrm{B,I:D}}{1-\eta_\mathrm{I:D}},\frac{\bar{n}_\mathrm{B,I:A}}{1-\eta_\mathrm{I:A}}\}$. The background noise mean photon number vector for Eve's detectors is $\bar{n}_{\mathrm{B},\mathfrak{E}}=\{ \frac{\bar{n}_{\mathrm{B},\mathfrak{E}:\mathrm{\tilde{h}}}}{1-\eta_{\mathfrak{E}:\mathrm{\tilde{h}}}},\frac{\bar{n}_{\mathrm{B},\mathfrak{E}:\mathrm{\tilde{v}}}}{1-\eta_{\mathfrak{E}:\mathrm{\tilde{v}}}},\frac{\bar{n}_{\mathrm{B},\mathfrak{E}:\mathrm{\tilde{d}}}}{1-\eta_{\mathfrak{E}:\mathrm{\tilde{d}}}},\frac{\bar{n}_{\mathrm{B},\mathfrak{E}:\mathrm{\tilde{a}}}}{1-\eta_{\mathfrak{E}:\mathrm{\tilde{a}}}}\}$. The background noise mean photon number vector for Alice's signal detectors when her untampered light is incident is $\bar{n}_\mathrm{B,S}=\{ \frac{\bar{n}_\mathrm{B,S:H}}{1-\eta_\mathrm{S:H}\xi},\frac{\bar{n}_\mathrm{B,S:V}}{1-\eta_\mathrm{S:V}\xi},\frac{\bar{n}_\mathrm{B,S:D}}{1-\eta_\mathrm{S:D}\xi},\frac{\bar{n}_\mathrm{B,S:A}}{1-\eta_\mathrm{S:A}\xi}\}$. Lastly, the background noise mean photon number vector for Alice's signal detectors when Eve's resent light is incident is $\bar{n}_{\mathrm{B,S}(\mathfrak{E})}=\{ \frac{\bar{n}_\mathrm{B,S:H}}{1-\eta_\mathrm{S:H}\xi_\mathfrak{E}},\frac{\bar{n}_\mathrm{B,S:V}}{1-\eta_\mathrm{S:V}\xi_\mathfrak{E}},\frac{\bar{n}_\mathrm{B,S:D}}{1-\eta_\mathrm{S:D}\xi_\mathfrak{E}},\frac{\bar{n}_\mathrm{B,S:A}}{1-\eta_\mathrm{S:A}\xi_\mathfrak{E}}\}$.

There are also the parameters which specify the delays (in time-bins) for the temporal spoofing scenario we consider, as detailed in Sect.~\ref{sect:rangefinding}. There is the delay from idler detectors to Alice's signal detectors $\delta_{\mathrm{I\to S}}$, the delay from idler detectors to Eve's detectors $\delta_{\mathrm{I\to}\mathfrak{ E}}$ and the delay from Eve's detectors to Alice's signal detectors $\delta_{\mathfrak{E}\mathrm{\to S}}$. However, as Alice does not know the specific delays which underlie the system we must define the delays they use for the two channels which they monitor. The real channel delay, from an idler click, is $\delta_\mathfrak{R}$ and the false channel delay, from an idler click is $\delta_\mathfrak{F}$. For the system considered here this relates to the aforementioned delay parameters as such $\delta_\mathfrak{R}=\delta_{\mathrm{I\to S}}$ and $\delta_\mathfrak{F}=\delta_{\mathrm{I\to}\mathfrak{ E}}+\delta_{\mathfrak{E}\mathrm{\to S}}$.

We can reduce the complexity of the event tree in our system by making some reasonable parameter assumptions. The possible combination of measurements from, for example, idler to Eve to Alice is a branch of the event tree. We assume that the idler background noise is $\bar{n}_\mathrm{B,I}\ll \mathbf{1}$, as the idler arm can easily be shielded. This assumption means that the branch of the system event tree where Alice measures a double idler click (both detectors for a basis registering a click) is negligible and so this branch can be neglected. We place a stronger assumption on the idler detector background noise in Sect.~\ref{sect:error_analysis}, as this avoids the possibility of registering a wrong coincidence click due to a false measurement at the idler arm. Another assumption is that Alice's signal mean photon number is $\bar{n}\ll 1$ due to the desire to avoid multi-photon contributions. This assumption means the rate of idler clicks is low and hence reduces the likelihood of polluted time-bins which complicate Alice's detection statistics. We define a polluted time-bin as when Eve resends light from a prior pulse of Alice's light or noise that they measure onto a channel which is populated by a subsequent pulse of Alice's light. For example, if Alice's first pulse defines a (first) real and false channel. Where Eve sends light to the first false channel. Then, Alice's second pulse defines a (second) real and false channel. If the second real channel overlaps with the first false channel we would refer to this as a polluted time-bin. The assumption that there is a limited number of polluted time-bins reduces the event tree complexity, as it means the statistics for Alice's two signal channels (real and false) are not dependent upon previous pulses of Alice's light or noise (that Eve measured) from earlier. We also assume that the background noise mean photon numbers for Eve's signal detectors are $\mathbf{\bar{n}_{\mathrm{B},\mathfrak{E}}}\ll 1$, to ensure that double clicks are negligible for Eve's detectors. The number of polluted time-bins is also limited thanks to this assumption.

The parameters used in Fig.~\ref{fig:relativebasischoice} are as follows $\eta_\mathrm{I}=\mathbf{0.99999}$, $\eta_\mathfrak{E}=\{0.799,0.099,0.009,0.989\}$, $\eta_\mathrm{S}=\{0.201,0.56,0.07,0.28\}$, $\bar{n}=4.975\times 10^{-3}$, $\xi_\mathfrak{E}=0.1$, $\xi=0.05$, $p_\mathfrak{R}=0$, $p_\mathfrak{F}=p$, $\bar{n}_\mathrm{B,I}=\mathbf{0.001}$, $\bar{n}_{\mathrm{B},\mathfrak{E}}=\mathbf{0.001}$, $\bar{n}_\mathrm{B,S}=\{0.025,0.0257,0.0251,0.0253\}$ and $\bar{n}_{\mathrm{B,S}(\mathfrak{E})}=\{0.0255,0.0265,0.0252,0.0257\}$.

The parameters used in Fig.~\ref{fig:SNRfig} and Fig.~\ref{fig:error_threshold} are as follows $\eta_\mathrm{I}=\mathbf{0.99999}$, $\eta_\mathfrak{E}=\mathbf{0.99999}$, $\eta_\mathrm{S}=\mathbf{0.9}$, $\bar{n}=4.975\times 10^{-3}$, $\bar{n}_\alpha=0.01$, $\xi_\mathfrak{E}=0.4$, $\xi=0.7$, $p_\mathfrak{R}=0$, $p_\mathfrak{F}=p$, $\bar{n}_\mathrm{B,I}=\mathbf{0.0001}$, $\bar{n}_{\mathrm{B},\mathfrak{E}}=\mathbf{0.0001}$, $\bar{n}_\mathrm{B,S}=\mathbf{0.375}$ and $\bar{n}_{\mathrm{B,S}(\mathfrak{E})}=\mathbf{0.648}$.

The parameters used in Fig.~\ref{fig:expectation_dists} and Fig.~\ref{fig:expectation_hist} are as follows. $\eta_\mathrm{I}=\mathbf{0.99999}$, $\eta_\mathfrak{E}=\mathbf{0.99999}$, $\eta_\mathrm{S}=\mathbf{0.7}$, $\bar{n}=4.975\times 10^{-3}$, $\bar{n}_\alpha=0.01$ ,$\xi_\mathfrak{E}=0.4$, $\xi=0.05$, $p_\mathfrak{R}=0$, $p_\mathfrak{F}=p$, $\bar{n}_\mathrm{B,I}=\mathbf{0.0001}$, $\bar{n}_{\mathrm{B},\mathfrak{E}}=\mathbf{0.0001}$, $\bar{n}_\mathrm{B,S}=\mathbf{0.259}$, $\bar{n}_{\mathrm{B,S}(\mathfrak{E})}=\mathbf{0.347}$, $\delta_\mathrm{I\to S}=2$, $\delta_{\mathrm{I}\to \mathfrak{E}}=1$, and $\delta_{\mathfrak{E}\to\mathrm{S}}=2$.

\subsection{Components}\label{appendix:components}
This subsection defines some components of the click probabilities for the QI system. The single-mode thermal state (which we use to model the background noise) is \begin{align}
    \hat{\rho}^\mathrm{B}_\mathcal{Q}(\bar{n}_{\mathrm{B}})&=\sum^\infty_{n=0}\mathrm{P}(n,\bar{n}_{\mathrm{B}})\vert n\rangle\langle n \vert_\mathcal{Q},\nonumber \\ &=\sum^\infty_{n=0}\frac{\bar{n}_{\mathrm{B}}^n}{(\bar{n}_\mathrm{B}+1)^{n+1}}\vert n \rangle\langle n \vert_\mathcal{Q},
\end{align}where $n$ is the photon number, $\mathcal{Q}$ is the mode and $\bar{n}_\mathrm{B}$ is the corresponding mean photon number. 

The following components are stated in general terms, such that they can have the relevant parameters substituted in for the click probability derivations. For example, we know which parameters to substitute in based off the mode labelling and the particular click probability we are calculating. The beamsplitter transmission parameter is related to the reflectivity parameter as such $\vert t\vert^2=1-\vert r \vert^2$. The first component is \begin{widetext}
    
\begin{align}
    \hat{\phi}(\bar{n}_{\mathrm{B}:\mathcal{Q}},\vert t_\mathcal{Q}\vert^2,\bar{n}_{\mathrm{B}:\mathcal{W}},\vert t_\mathcal{W}\vert^2)&=\sum^\infty_{q=0}\mathrm{P}(q,\bar{n}_{\mathrm{B}:\mathcal{Q}})\sum^{q}_{z=0}\binom{q}{z}\vert r_\mathcal{Q}\vert^{2z}\vert t_\mathcal{Q}\vert^{2(q-z)}\times\nonumber \\ &\times\biggl(\vert t_\mathcal{Q}\vert^2(z+1)\vert z+1\rangle\langle z+1\vert_\mathcal{Q}+\vert r_\mathcal{Q}\vert^2(q-z+1)\vert z\rangle\langle z \vert_\mathcal{Q}\biggr)\otimes\hat{\rho}^\mathrm{B}_\mathcal{W}(\vert r_\mathcal{W}\vert^2\bar{n}_{\mathrm{B},\mathcal{W}}),\label{eq:generalised_phi}
\end{align} where $\mathcal{Q}$ and $\mathcal{W}$ are mode labellings for the two elements in the chosen basis. This component (Eq.~\ref{eq:generalised_phi}) describes the mixing of a single-photon with a thermal state in a tensor product with a thermal state in another mode. Throughout these appendices, the order of the mode labelling in the argument signifies which contains signal light mixed with background noise and which only contains background noise. In Eq.~\ref{eq:generalised_phi} we can see that mode $\mathcal{Q}$ contains signal light mixed with background noise and mode $\mathcal{W}$ only contains background noise.

We define the normalisation factor required due to the inclusion of loss and noise on the idler beam (derived from Eq.~\ref{eq:state}) as \begin{equation}
    \mathcal{N}_\mathrm{I}=\mathrm{Tr}\biggl(c_0 \hat{\rho}^\mathrm{B}_\mathcal{Q}(\vert r_\mathcal{Q}\vert^2\bar{n}_{\mathrm{B},\mathcal{Q}})\otimes \hat{\rho}^\mathrm{B}_\mathcal{W}(\vert r_\mathcal{W}\vert^2\bar{n}_{\mathrm{B},\mathcal{W}})+c_1\bigl(\hat{\phi}(\bar{n}_{\mathrm{B}:\mathcal{Q}},\eta_{\mathcal{Q}},\bar{n}_{\mathrm{B}:\mathcal{W}},\eta_{\mathcal{W}})+\hat{\phi}(\bar{n}_{\mathrm{B}:\mathcal{W}},\eta_{\mathcal{W}},\bar{n}_{\mathrm{B}:\mathcal{Q}},\eta_{\mathcal{Q}})\bigr)\biggr).
\end{equation}  As $\mathcal{Q}$ and $\mathcal{W}$ signifies the idler beam basis elements here, then $\mathcal{J}$ and $\mathcal{K}$ signifies the signal mode basis elements. By convention $\mathcal{J}$ is the correlated mode to $\mathcal{Q}$ when $\theta_\mathrm{T}=0$. In general, however, the correlation of idler mode to signal mode is a function of $\theta_\mathrm{T}$ according to Eq.~\ref{eq:basis_relations}. Hence, we can write Alice's state of light after the inclusion of noise and loss on the idler beam as \begin{align}
    \hat{\rho}_\mathrm{NORM}&=\frac{1}{\mathcal{N}_\mathrm{I}}\biggl(c_0\hat{\rho}^\mathrm{B}_\mathcal{Q}(\vert r_\mathcal{Q}\vert^2\bar{n}_{\mathrm{B},\mathcal{Q}})\otimes \hat{\rho}^\mathrm{B}_\mathcal{W}(\vert r_\mathcal{W}\vert^2\bar{n}_{\mathrm{B},\mathcal{W}})\otimes \vert 0\rangle\langle 0\vert_\mathcal{J} \otimes \vert 0\rangle\langle 0\vert_\mathcal{K}+ \nonumber \\ &+c_1\bigl(\hat{\phi}(\bar{n}_{\mathrm{B}:\mathcal{Q}},\eta_{\mathcal{Q}},\bar{n}_{\mathrm{B}:\mathcal{W}},\eta_{\mathcal{W}})\otimes(\mathrm{cos}^2(\theta_\mathrm{T}) \vert 1\rangle\langle 1\vert_\mathcal{J} \otimes \vert 0\rangle\langle 0\vert_\mathcal{K}+\mathrm{sin}^2(\theta_\mathrm{T})\vert 0\rangle\langle 0\vert_\mathcal{J} \otimes \vert 1\rangle\langle 1\vert_\mathcal{K})+\nonumber\\&+\hat{\phi}(\bar{n}_{\mathrm{B}:\mathcal{W}},\eta_{\mathcal{W}},\bar{n}_{\mathrm{B}:\mathcal{Q}},\eta_{\mathcal{Q}})\otimes(\mathrm{sin}^2(\theta_\mathrm{T}) \vert 1\rangle\langle 1\vert_\mathcal{J} \otimes \vert 0\rangle\langle 0\vert_\mathcal{K}+\mathrm{cos}^2(\theta_\mathrm{T})\vert 0\rangle\langle 0\vert_\mathcal{J} \otimes \vert 1\rangle\langle 1\vert_\mathcal{K})\bigr)\biggr).
\end{align}
 If there is a click for the $\mathcal{Q}$ mode and no-click for the $\mathcal{W}$ mode the vacuum contribution component (the sub-state associated with coefficient $c_0$) is \begin{equation}
    \mathrm{vac}(\mathcal{Q})=\left(1-\frac{1}{1+\vert r_\mathcal{Q}\vert^2 \bar{n}_{\mathrm{B}:\mathcal{Q}}}\right)\left(\frac{1}{1+\vert r_\mathcal{W}\vert^2 \bar{n}_{\mathrm{B}:\mathcal{W}}}\right).
\end{equation}If there is a click for the $\mathcal{Q}$ mode and no-click for the $\mathcal{W}$ mode the single-photon contribution components (associated with coefficient $c_1$) are \begin{align}
    a_{\mathcal{Q}\vert 1}&=\mathrm{Tr}\left(\hat{\phi}(\bar{n}_{\mathrm{B}:\mathcal{Q}},\vert t_\mathcal{Q}\vert^2,\bar{n}_{\mathrm{B}:\mathcal{W}},\vert t_\mathcal{W}\vert^2)(\hat{1}_\mathcal{Q}-\vert 0 \rangle\langle 0\vert_\mathcal{Q})\otimes \vert 0 \rangle \langle 0 \vert_\mathcal{W}\right),\nonumber \\ &=\sum^\infty_{q=0}\mathrm{P}(q,\bar{n}_{\mathrm{B}:\mathcal{Q}})\biggl[\sum^{q}_{z=0}\binom{q}{z}\vert r_\mathcal{Q}\vert^{2z}\vert t_\mathcal{Q}\vert^{2(q-z)}\biggl(\vert t_\mathcal{Q}\vert^2(z+1)+\vert r_\mathcal{Q}\vert^2(q-z+1)\biggr)-\vert t_\mathcal{Q}\vert^{2q}\vert r_\mathcal{Q}\vert^2(q+1)\biggr]\times\nonumber \\ &\times \left(\frac{1}{1+\vert r_\mathcal{W}\vert^2 \bar{n}_{\mathrm{B}:\mathcal{W}}}\right),
\label{eq:a1}\end{align} and
\begin{align}
    a_{\mathcal{W}\vert 0}&=\mathrm{Tr}\left(\hat{\phi}(\bar{n}_{\mathrm{B}:\mathcal{W}},\vert t_\mathcal{W}\vert^2,\bar{n}_{\mathrm{B}:\mathcal{Q}},\vert t_\mathcal{Q}\vert^2)(\hat{1}_\mathcal{Q}-\vert 0 \rangle\langle 0\vert_\mathcal{Q})\otimes \vert 0 \rangle \langle 0 \vert_\mathcal{W}\right),\nonumber \\ &=\sum^\infty_{q=0}\mathrm{P}(q,\bar{n}_{\mathrm{B}:\mathcal{W}})\biggl[\vert t_\mathcal{W}\vert^{2q}\vert r_\mathcal{W}\vert^2(q+1)\biggr]\times \left(1-\frac{1}{1+\vert r_\mathcal{Q}\vert^2 \bar{n}_{\mathrm{B}:\mathcal{Q}}}\right).
\label{eq:a0}\end{align} If there is a click for both $\mathcal{Q}$ and $\mathcal{W}$ the vacuum contribution is \begin{equation}
    \mathrm{vac}(\mathrm{Q},\mathrm{W})=\left(1-\frac{1}{1+\vert r_\mathcal{Q}\vert^2 \bar{n}_{\mathrm{B}:\mathcal{Q}}}\right)\left(1-\frac{1}{1+\vert r_\mathcal{W}\vert^2 \bar{n}_{\mathrm{B}:\mathcal{W}}}\right),
\end{equation} If there is a click for both $\mathcal{Q}$ and $\mathcal{W}$ the single-photon contribution component is \begin{align}
    a_{\mathcal{Q,W}\vert 1}&=\mathrm{Tr}\left(\hat{\phi}(\bar{n}_{\mathrm{B}:\mathcal{Q}},\vert t_\mathcal{Q}\vert^2,\bar{n}_{\mathrm{B}:\mathcal{W}},\vert t_\mathcal{W}\vert^2)(\hat{1}_\mathcal{Q}-\vert 0 \rangle\langle 0\vert_\mathcal{Q})\otimes(\hat{1}_\mathcal{W}- \vert 0 \rangle \langle 0 \vert_\mathcal{W})\right),\nonumber \\ &=\sum^\infty_{q=0}\mathrm{P}(q,\bar{n}_{\mathrm{B}:\mathcal{Q}})\biggl[\sum^{q}_{z=0}\binom{q}{z}\vert r_\mathcal{Q}\vert^{2z}\vert t_\mathcal{Q}\vert^{2(q-z)}\biggl(\vert t_\mathcal{Q}\vert^2(z+1)+\vert r_\mathcal{Q}\vert^2(q-z+1)\biggr)-\vert t_\mathcal{Q}\vert^{2q}\vert r_\mathcal{Q}\vert^2(q+1)\biggr]\times\nonumber \\ &\times \left(1-\frac{1}{1+\vert r_\mathcal{W}\vert^2 \bar{n}_{\mathrm{B}:\mathcal{W}}}\right).
\end{align}
We can see that these components are in terms of the two elements of a certain basis. These component definitions are useful for the sake of brevity in our derivations. The no-click component (when both mode $\mathcal{Q}$ and $\mathcal{W}$ does not register a click) for the noise-only contribution is \begin{equation}
    \mathrm{vac}_0(\mathcal{Q},\mathcal{W})=\biggl(\frac{1}{1+\vert r_\mathcal{Q}\vert^2\bar{n}_\mathrm{B,\mathcal{Q}}}\biggr)\biggl(\frac{1}{1+\vert r_\mathcal{W}\vert^2\bar{n}_\mathrm{B,\mathcal{W}}}\biggr).
\end{equation} The no-click component for the single-photon contribution, where mode $\mathcal{Q}$ contains Alice's single photon and noise and mode $\mathcal{W}$ contains only noise is
    \begin{equation}
      a_\mathrm{\mathcal{Q,W}\vert 0}=\frac{1}{1+\vert r_\mathcal{W}\vert^2\bar{n}_\mathrm{B,\mathcal{W}}}\sum^\infty_{m=0}\frac{\bar{n}_\mathrm{B,\mathcal{Q}}^m}{(\bar{n}_\mathrm{B,\mathcal{Q}}+1)^{m+1}}\vert t_\mathcal{Q}\vert^{2m} \vert r_\mathcal{Q}\vert^2(m+1).
    \end{equation}
\end{widetext}

\subsection{Idler}
In the regime where the signal mean photon number $\bar{n}\ll 1$ and the idler detector background noise mean photon number $\mathbf{\bar{n}_\mathrm{B,I}}\ll\mathbf{1}$ double coincidence idler clicks are negligible and so we focus only on events where there is a click for one mode and no-click for another.

If there is a click for $\mathcal{Q}$ and no-click for $\mathcal{W}$ the idler detector sub-system click probability is 
\begin{align}
    \mathrm{Pr}_\mathrm{I:\mathcal{Q}}&=\mathrm{Tr}\left(\hat{\rho}_\mathrm{NORM}(\hat{1}_\mathcal{Q}-\vert 0\rangle \langle 0\vert_\mathcal{Q})\otimes \vert 0 \rangle \langle 0\vert_\mathcal{W}\right),\nonumber \\ 
    &=\frac{c_0}{\mathcal{N}_\mathrm{I}}\left(\mathrm{vac}(\mathcal{Q})\right)+\frac{c_1}{\mathcal{N}_\mathrm{I}}\left(a_{\mathcal{Q}\vert 1}+a_{\mathcal{W}\vert 0}\right).
\end{align} 
\begin{widetext}
Measurement of the idler modes conditions the state which will be sent towards Eve or Alice. Of course, Eve does not have access to the idler mode measurement, but she could still receive this idler conditioned state.  The idler click (click for $\mathcal{Q}$ and no-click for $\mathcal{W}$) conditioned state is \begin{align}
    \hat{\rho}_{\mathrm{I}:\mathcal{Q}\vert 1}&=\frac{\mathrm{Tr}_\mathrm{I}\left(\hat{\rho}_\mathrm{NORM}(\hat{1}_\mathcal{Q}-\vert 0\rangle \langle 0\vert_\mathcal{Q})\otimes \vert 0 \rangle \langle 0\vert_\mathcal{W}\right)}{\mathrm{Pr}_\mathrm{I:\mathcal{Q}}},\nonumber \\ &=\frac{1}{\mathrm{Pr}_\mathrm{I:\mathcal{Q}}}\biggl(\frac{c_0}{\mathcal{N}_\mathrm{I}}\left(\mathrm{vac}(\mathcal{Q})\vert 0\rangle\langle 0\vert_\mathcal{J}\otimes \vert 0\rangle\langle 0\vert_\mathcal{K}\right)+\frac{c_1}{\mathcal{N}_\mathrm{I}}\left(q_2(\theta_\mathrm{T})\vert 1\rangle\langle 1\vert_\mathcal{J}\otimes\vert 0\rangle\langle 0\vert_\mathcal{K}+q_1(\theta_\mathrm{T})\vert 0\rangle\langle 0\vert_\mathcal{J}\otimes\vert 1\rangle\langle 1\vert_\mathcal{K}\right)\biggr),\label{eq:idler_click_cond}
\end{align} where we define \begin{align}
    q_1(\theta_\mathrm{T})&=a_{\mathcal{Q}\vert 1}\mathrm{sin}^2(\theta_\mathrm{T})+a_{\mathcal{W}\vert 0}\mathrm{cos}^2(\theta_\mathrm{T}),\\
    q_2(\theta_\mathrm{T})&=a_{\mathcal{Q}\vert 1}\mathrm{cos}^2(\theta_\mathrm{T})+a_{\mathcal{W}\vert 0}\mathrm{sin}^2(\theta_\mathrm{T}).
\end{align} 
\end{widetext} \subsection{Signal}
When the idler conditioned state (Eq.~\ref{eq:idler_click_cond}) is at the signal detector sub-system it has its basis elements separated and noise for each basis element mode included, we denote this state as $\hat{\rho}_\mathrm{S\vert I:\mathcal{Q}}^\mathrm{UNNORM}$. We normalise this state with normalisation factor $\mathcal{N}_\mathrm{S\vert I:\mathcal{Q}}=\mathrm{Tr}\left(\hat{\rho}_\mathrm{S\vert I:\mathcal{Q}}^\mathrm{UNNORM}\right)$. Where this normalisation factor is \begin{widetext}\begin{align}
  \mathcal{N}_\mathrm{S\vert I:\mathcal{Q}}&=\frac{1}{\mathrm{Pr}_\mathrm{I:\mathcal{Q}}\mathcal{N}_\mathrm{I}}\biggl(c_0\mathrm{vac}(\mathcal{Q})\hat{\rho}^\mathrm{B}_\mathcal{J}(\vert r_\mathcal{J}\vert^2\bar{n}_\mathrm{B:\mathcal{J}})\otimes \hat{\rho}^\mathrm{B}_\mathcal{K}(\vert r_\mathcal{K}\vert^2\bar{n}_\mathrm{B:\mathcal{K}})+\nonumber\\&+c_1\bigl( q_2(\theta_\mathrm{T})\hat{\phi}(\bar{n}_{\mathrm{B}:\mathcal{J}},\vert t_\mathcal{J}\vert^2,\bar{n}_{\mathrm{B}:\mathcal{K}},\vert t_\mathcal{K}\vert^2)+q_1(\theta_\mathrm{T})\hat{\phi}(\bar{n}_{\mathrm{B}:\mathcal{K}},\vert t_\mathcal{K}\vert^2,\bar{n}_{\mathrm{B}:\mathcal{J}},\vert t_\mathcal{J}\vert^2)\bigr)\biggr).
\end{align}
The normalised state incident upon the signal detector sub-system is 
    \begin{align}
        \hat{\rho}_\mathrm{S\vert I:\mathcal{Q}}&=\frac{\hat{\rho}_\mathrm{S\vert I:\mathcal{Q}}^\mathrm{UNNORM}}{\mathcal{N}_\mathrm{S\vert I:\mathcal{Q}}}, \nonumber\\ &=\frac{1}{\mathcal{N}_\mathrm{I}\mathrm{Pr}_\mathrm{I:\mathcal{Q}}\mathcal{N}_\mathrm{S\vert I:\mathcal{Q}}}\biggl(c_0\mathrm{vac}(\mathcal{Q})\hat{\rho}^\mathrm{B}_\mathcal{J}(\vert r_\mathcal{J}\vert^2\bar{n}_\mathrm{B:\mathcal{J}})\otimes \hat{\rho}^\mathrm{B}_\mathcal{K}(\vert r_\mathcal{K}\vert^2\bar{n}_\mathrm{B:\mathcal{K}})+\nonumber \\ &+c_1\bigl( q_2(\theta_\mathrm{T})\hat{\phi}(\bar{n}_{\mathrm{B}:\mathcal{J}},\vert t_\mathcal{J}\vert^2,\bar{n}_{\mathrm{B}:\mathcal{K}},\vert t_\mathcal{K}\vert^2)+q_1(\theta_\mathrm{T})\hat{\phi}(\bar{n}_{\mathrm{B}:\mathcal{K}},\vert t_\mathcal{K}\vert^2,\bar{n}_{\mathrm{B}:\mathcal{J}},\vert t_\mathcal{J}\vert^2)\bigr)\biggr).
    \end{align} The following click probabilities can apply to Eve's signal detector or Alice's signal detector (if it is her untampered light incident). Hence, the click probability when signal detector for mode $\mathcal{J}$ clicks and $\mathcal{K}$ does not, when conditioned by an idler detector for mode $\mathcal{Q}$ click is \begin{align}
        \mathrm{Pr}_\mathrm{S:\mathcal{J}\vert I:\mathcal{Q}}(\theta_\mathrm{T})&=\mathrm{Tr}\left(\hat{\rho}_\mathrm{S\vert I:\mathcal{Q}}(\hat{1}_\mathcal{J}-\vert 0\rangle\langle 0\vert_\mathcal{J})\otimes \vert 0 \rangle\langle 0\vert_\mathcal{K}\right),\nonumber \\ &=\frac{1}{\mathcal{N}_\mathrm{I}\mathrm{Pr}_\mathrm{I:\mathcal{Q}}\mathcal{N}_\mathrm{S\vert I:\mathcal{Q}}}\biggl(c_0 \mathrm{vac}(\mathcal{Q}) \mathrm{vac}(\mathcal{J})+c_1\bigl( q_2(\theta_\mathrm{T})a_{\mathcal{J}\vert 1}+q_1(\theta_\mathrm{T})a_{\mathcal{K}\vert 0}\bigr)\biggr).
    \end{align} The click probability when signal detector for mode $\mathcal{K}$ clicks and $\mathcal{J}$ does not, when conditioned by an idler detector for mode $\mathcal{Q}$ click is \begin{align}
        \mathrm{Pr}_\mathrm{S:\mathcal{K}\vert I:\mathcal{Q}}(\theta_\mathrm{T})&=\mathrm{Tr}\left(\hat{\rho}_\mathrm{S\vert I:\mathcal{Q}}(\hat{1}_\mathcal{K}-\vert 0\rangle\langle 0\vert_\mathcal{K})\otimes \vert 0 \rangle\langle 0\vert_\mathcal{J}\right),\nonumber \\ &=\frac{1}{\mathcal{N}_\mathrm{I}\mathrm{Pr}_\mathrm{I:\mathcal{Q}}\mathcal{N}_\mathrm{S\vert I:\mathcal{Q}}}\biggl(c_0 \mathrm{vac}(\mathcal{Q}) \mathrm{vac}(\mathcal{K})+c_1\bigl( q_2(\theta_\mathrm{T})a_{\mathcal{J}\vert 0}+q_1(\theta_\mathrm{T})a_{\mathcal{K}\vert 1}\bigr)\biggr).
    \end{align} The click probability when both the signal detector for mode $\mathcal{K}$ and $\mathcal{J}$ clicks, when conditioned by an idler detector for mode $\mathcal{Q}$ click is 
    \begin{align}
        \mathrm{Pr}_\mathrm{S:\mathcal{J,K}\vert I:\mathcal{Q}}(\theta_\mathrm{T})&=\mathrm{Tr}\left(\hat{\rho}_\mathrm{S\vert I:\mathcal{Q}}(\hat{1}_\mathcal{J}-\vert 0\rangle\langle 0\vert_\mathcal{J})\otimes(\hat{1}_\mathcal{K} -\vert 0 \rangle\langle 0\vert_\mathcal{K})\right),\nonumber \\ &=\frac{1}{\mathcal{N}_\mathrm{I}\mathrm{Pr}_\mathrm{I:\mathcal{Q}}\mathcal{N}_\mathrm{S\vert I:\mathcal{Q}}}\biggl(c_0 \mathrm{vac}(\mathcal{Q}) \mathrm{vac}(\mathcal{J,K})+c_1\bigl( q_2(\theta_\mathrm{T})a_{\mathcal{J,K}\vert 1}+q_1(\theta_\mathrm{T})a_{\mathcal{K,J}\vert 1}\bigr)\biggr).
    \end{align} 
    The click probability for Eve when neither signal detector for the modes in the set $\mathcal{L}\in \{\mathcal{J},\mathcal{K}\}$ clicks is \begin{align}
   \mathrm{Pr}_\mathrm{S0:\mathcal{L}\vert I:\mathcal{Q}}(\theta_\mathrm{T})&= \mathrm{Tr}\left(\hat{\rho}_\mathrm{S\vert I:\mathcal{Q}}\vert 0\rangle\langle 0\vert_{\mathcal{J}}\otimes\vert 0 \rangle\langle 0\vert_\mathcal{K}\right),\nonumber\\ &=\frac{1}{\mathcal{N}_\mathrm{I}\mathrm{Pr}_\mathrm{I:\mathcal{Q}}\mathcal{N}_\mathrm{S\vert I:\mathcal{Q}}}\biggl(c_0 \mathrm{vac}(\mathcal{Q})\mathrm{vac}_0(\mathcal{J,K})+c_1\bigl(q_2(\theta_\mathrm{T})a_{\mathcal{J,K}\vert 0}+q_1(\theta_\mathrm{T})a_{\mathcal{K,J}\vert 0}\bigr)\biggr).
\end{align} The corresponding click probability for Alice is not considered, as Alice does not include her signal detector no-click events for her detection statistics, whereas the situation of Eve not resending a photon is relevant to what Alice (does and does not) receive.

\end{widetext} As described earlier in this paper, Eve generates a single-photon corresponding to what she measures and she then sends this single-photon to Alice's signal detectors. We denote this as $\mathcal{J}_\mathfrak{E}$. There is the unnormalised state after the basis elements have been separated and noise included at Alice's signal detector sub-system $\hat{\rho}_\mathrm{S:\mathcal{J}\vert \mathcal{J}_\mathfrak{E}}^{\mathrm{UNNORM}}$, which is normalised by $\mathcal{N}_\mathfrak{E}=\mathrm{Tr}\left(\hat{\rho}_\mathrm{S:\mathcal{J}\vert \mathcal{J}_\mathfrak{E}}^{\mathrm{UNNORM}}\right)$. Where this normalisation factor is defined as \begin{widetext} \begin{equation}
    \mathcal{N}_\mathfrak{E}=\mathrm{cos}^2(\theta_\mathrm{T})\hat{\phi}(\bar{n}_{\mathrm{B}:\mathcal{J}},\vert t_\mathcal{J}\vert^2,\bar{n}_{\mathrm{B}:\mathcal{K}},\vert t_\mathcal{K}\vert^2)+\mathrm{sin}^2(\theta_\mathrm{T})\hat{\phi}(\bar{n}_{\mathrm{B}:\mathcal{K}},\vert t_\mathcal{K}\vert^2,\bar{n}_{\mathrm{B}:\mathcal{J}},\vert t_\mathcal{J}\vert^2).
\end{equation}The resultant normalised state (where we have once again excluded off-diagonal terms) is \begin{align}
    \hat{\rho}_\mathrm{S:\mathcal{J}\vert E: \mathcal{J}_\mathfrak{E}}&=\frac{\hat{\rho}_\mathrm{S:\mathcal{J}\vert E:\mathcal{J}_\mathfrak{E}}^{\mathrm{UNNORM}}}{\mathcal{N}_\mathfrak{E}},\nonumber \\ &=\frac{1}{\mathcal{N}_\mathfrak{E}}\left(\mathrm{cos}^2(\theta_\mathrm{T})\hat{\phi}(\bar{n}_{\mathrm{B}:\mathcal{J}},\vert t_\mathcal{J}\vert^2,\bar{n}_{\mathrm{B}:\mathcal{K}},\vert t_\mathcal{K}\vert^2)+\mathrm{sin}^2(\theta_\mathrm{T})\hat{\phi}(\bar{n}_{\mathrm{B}:\mathcal{K}},\vert t_\mathcal{K}\vert^2,\bar{n}_{\mathrm{B}:\mathcal{J}},\vert t_\mathcal{J}\vert^2)\right).\label{eq:eve_resent}
\end{align} The state (Eq.~\ref{eq:eve_resent}) thus depends on the relative basis angle and basis choice of Eve, where $\mathcal{J}_\mathfrak{E}$ corresponds to the mode $\mathcal{J}$ (as seen by Alice) when $\theta_\mathrm{T}=0$. The following click probabilities are for when Eve's resent light is incident upon Alice's signal detector subsystem. Hence, the click probability when the signal detector for mode $\mathcal{J}$ clicks and $\mathcal{K}$ does not click (when a single-photon $\mathcal{J}_\mathfrak{E}$ is sent) is\begin{align}
    \mathrm{Pr}_\mathrm{S:\mathcal{J}\vert E: \mathcal{J}_\mathfrak{E}}(\theta_\mathrm{T})&=\mathrm{Tr}\left(\hat{\rho}_\mathrm{S:\mathcal{J}\vert E:\mathcal{J}_\mathfrak{E}}(\hat{1}_{\mathcal{J}}-\vert 0\rangle\langle 0\vert_{\mathcal{J}})\otimes\vert 0 \rangle\langle 0\vert_\mathcal{K}\right), \nonumber\\
    &=\frac{1}{\mathcal{N}_\mathfrak{E}}\left(\mathrm{cos}^2(\theta_\mathrm{T})a_{\mathcal{J}\vert 1}+\mathrm{sin}^2(\theta_\mathrm{T})a_{\mathcal{K}\vert 0}\right).
\end{align} The click probability when the signal detector for $\mathcal{K}$ clicks and $\mathcal{J}$ does not click is  \begin{align}
    \mathrm{Pr}_\mathrm{S:\mathcal{K}\vert E:\mathcal{J}_\mathfrak{E}}(\theta_\mathrm{T})&=\mathrm{Tr}\left(\hat{\rho}_\mathrm{S:\mathcal{J}\vert E:\mathcal{J}_\mathfrak{E}}(\hat{1}_{\mathcal{K}}-\vert 0\rangle\langle 0\vert_{\mathcal{K}})\otimes\vert 0 \rangle\langle 0\vert_\mathcal{J}\right),\nonumber \\
    &=\frac{1}{\mathcal{N}_\mathfrak{E}}\left(\mathrm{cos}^2(\theta_\mathrm{T})a_{\mathcal{J}\vert 0}+\mathrm{sin}^2(\theta_\mathrm{T})a_{\mathcal{K}\vert 1}\right).
\end{align}The click probability when the signal detector for both $\mathcal{J}$ and $\mathcal{K}$ clicks is
 \begin{align}
    \mathrm{Pr}_\mathrm{S:\mathcal{J,K}\vert E:\mathcal{J}_\mathfrak{E}}(\theta_\mathrm{T})&=\mathrm{Tr}\left(\hat{\rho}_\mathrm{S:\mathcal{J}\vert E:\mathcal{J}_\mathfrak{E}}(\hat{1}_{\mathcal{K}}-\vert 0\rangle\langle 0\vert_{\mathcal{K}})\otimes(\hat{1}_\mathcal{J}-\vert 0 \rangle\langle 0\vert_\mathcal{J})\right),\nonumber \\
    &=\frac{1}{\mathcal{N}_\mathfrak{E}}\left(\mathrm{cos}^2(\theta_\mathrm{T})a_{\mathcal{J,K}\vert 1}+\mathrm{sin}^2(\theta_\mathrm{T})a_{\mathcal{K,J}\vert 1}\right).
\end{align} 

\end{widetext}
\subsection{Alice}
We define the untampered correct coincidence, wrong coincidence and double coincidence contributions from Alice in this section. The total angle is zero $\theta_\mathrm{T}=0$ for these contributions as there is no relative basis angle and Alice always chooses the correct basis (therefore there will not be a $\frac{\pi}{4}$ shift). Therefore, we omit the $\theta_\mathrm{T}$ notation for the contributions from Alice. The Alice contribution correct coincidence probability is
\begin{widetext}
\begin{equation}
    \mathrm{Pr}^\mathfrak{A}_{\mathrm{c}}=\frac{1}{2}\biggl(\sum^2_{\substack{i=1 \\ i\neq j}} \mathrm{Pr}_{\mathrm{I:\mathcal{X}_{+,i}}}\mathrm{Pr}_{\mathrm{S:\mathcal{X}_{+,j}}\vert\mathrm{I:\mathcal{X}_{+,i}}}
+\sum^2_{i=1}\mathrm{Pr}_{\mathrm{I:\mathcal{X}_{\times,i}}}\mathrm{Pr}_{\mathrm{S:\mathcal{X}_{\times,i}}\vert\mathrm{I:\mathcal{X}_{\times,i}}}\biggr).
\end{equation} The Alice contribution wrong coincidence probability is \begin{equation}
    \mathrm{Pr}^\mathfrak{A}_{\mathrm{w}}=\frac{1}{2}\biggl(\sum^2_{i=1} \mathrm{Pr}_{\mathrm{I:\mathcal{X}_{+,i}}}\mathrm{Pr}_{\mathrm{S:\mathcal{X}_{+,i}}\vert\mathrm{I:\mathcal{X}_{+,i}}}
+\sum^2_{\substack{i=1 \\ i\neq j}}\mathrm{Pr}_{\mathrm{I:\mathcal{X}_{\times,i}}}\mathrm{Pr}_{\mathrm{S:\mathcal{X}_{\times,j}}\vert\mathrm{I:\mathcal{X}_{\times,i}}}\biggr).
\end{equation}The Alice contribution double coincidence probability is \begin{equation}
    \mathrm{Pr}^\mathfrak{A}_{\mathrm{wc}}=\frac{1}{2}\biggl(\sum^2_{\substack{i=1 \\ i\neq j}} \mathrm{Pr}_{\mathrm{I:\mathcal{X}_{+,i}}}\mathrm{Pr}_{\mathrm{S:\mathcal{X}_{+,i},\mathcal{X}_{+,j}}\vert\mathrm{I:\mathcal{X}_{+,i}}}
+\sum^2_{\substack{i=1 \\ i\neq j}}\mathrm{Pr}_{\mathrm{I:\mathcal{X}_{\times,i}}}\mathrm{Pr}_{\mathrm{S:\mathcal{X}_{\times,i},\mathcal{X}_{\times,j}}\vert\mathrm{I:\mathcal{X}_{\times,i}}}\biggr).
\end{equation}
\end{widetext}
\subsection{Eve}
In this section we define the correct, wrong and double coincidence contributions from Eve upon Alice's signal detector. The Eve contribution correct coincidence probability is \begin{widetext}
    \begin{equation}\begin{split}
    \mathrm{Pr}^\mathfrak{E}_\mathrm{c}(\theta)&=\frac{1}{2}\sum^2_{\substack{i=1 \\ i\neq j}}\mathrm{Pr}_{\mathrm{I:\mathcal{X}_{+,i}}}\biggl[ r \biggl(\mathrm{Pr}_{\mathrm{S:\mathcal{X}_{+,j}}\vert\mathrm{E:\mathcal{Y}_{+,i}}}(\theta_\mathrm{T})\mathrm{Pr}_{\mathrm{S:\mathcal{Y}_{+,i}}\vert\mathrm{I:\mathcal{X}_{+,i}}}(\theta_\mathrm{T})+\mathrm{Pr}_{\mathrm{S:\mathcal{X}_{+,j}}\vert\mathrm{E:\mathcal{Y}_{+,j}}}(\theta_\mathrm{T})\mathrm{Pr}_{\mathrm{S:\mathcal{Y}_{+,j}}\vert\mathrm{I:\mathcal{X}_{+,i}}}(\theta_\mathrm{T})\biggr)+ \\ &+(1-r)\biggl(\mathrm{Pr}_{\mathrm{S:\mathcal{X}_{+,j}}\vert\mathrm{E:\mathcal{Y}_{\times,i}}}(\theta_\mathrm{T})\mathrm{Pr}_{\mathrm{S:\mathcal{Y}_{\times,i}}\vert\mathrm{I:\mathcal{X}_{+,i}}}(\theta_\mathrm{T})+\mathrm{Pr}_{\mathrm{S:\mathcal{X}_{+,j}}\vert\mathrm{E:\mathcal{X}_{\times,j}}}(\theta_\mathrm{T})\mathrm{Pr}_{\mathrm{S:\mathcal{Y}_{\times,j}}\vert\mathrm{I:\mathcal{X}_{+,i}}}(\theta_\mathrm{T})\biggr)\biggr]+ \\ &+\frac{1}{2}\sum^2_{\substack{i=1 \\ i\neq j}}\mathrm{Pr}_{\mathrm{I:\mathcal{X}_{\times,i}}}\biggl[ r \biggl(\mathrm{Pr}_{\mathrm{S:\mathcal{X}_{\times,i}}\vert\mathrm{E:\mathcal{Y}_{+,i}}}(\theta_\mathrm{T})\mathrm{Pr}_{\mathrm{S:\mathcal{Y}_{+,i}}\vert\mathrm{I:\mathcal{X}_{\times,i}}}(\theta_\mathrm{T})+\mathrm{Pr}_{\mathrm{S:\mathcal{X}_{\times,i}}\vert\mathrm{E:\mathcal{Y}_{+,j}}}(\theta_\mathrm{T})\mathrm{Pr}_{\mathrm{S:\mathcal{Y}_{+,j}}\vert\mathrm{I:\mathcal{X}_{\times,i}}}(\theta_\mathrm{T})\biggr)+ \\ &+(1-r)\biggl(\mathrm{Pr}_{\mathrm{S:\mathcal{X}_{\times,i}}\vert\mathrm{E:\mathcal{Y}_{\times,i}}}(\theta_\mathrm{T})\mathrm{Pr}_{\mathrm{S:\mathcal{Y}_{\times,i}}\vert\mathrm{I:\mathcal{X}_{\times,i}}}(\theta_\mathrm{T})+\mathrm{Pr}_{\mathrm{S:\mathcal{X}_{\times,i}}\vert\mathrm{E:\mathcal{X}_{\times,j}}}(\theta_\mathrm{T})\mathrm{Pr}_{\mathrm{S:\mathcal{Y}_{\times,j}}\vert\mathrm{I:\mathcal{X}_{\times,i}}}(\theta_\mathrm{T})\biggr)\biggr]+ \\
    &+\frac{1}{2}\sum^2_{\substack{i=1 \\ i\neq j}}\biggl[\mathrm{Pr}_{\mathrm{I:\mathcal{X}_{+,i}}}\mathrm{vac}(\mathcal{X}_{+,j})\biggl(r\mathrm{Pr}_{\mathrm{S0:\mathcal{Y}_+\vert I:\mathcal{X}_{+,i}}}(\theta_\mathrm{T})+(1-r)\mathrm{Pr}_{\mathrm{S0:\mathcal{Y}_\times\vert I:\mathcal{X}_{+,i}}}(\theta_\mathrm{T})\biggr)+\\ &+\mathrm{Pr}_{\mathrm{I:\mathcal{X}_{\times,i}}}\mathrm{vac}(\mathcal{X}_{\times,i})\biggl(r\mathrm{Pr}_{\mathrm{S0:\mathcal{Y}_+\vert I:\mathcal{X}_{\times,i}}}(\theta_\mathrm{T})+(1-r)\mathrm{Pr}_{\mathrm{S0:\mathcal{Y}_\times\vert I:\mathcal{X}_{\times,i}}}(\theta_\mathrm{T})\biggr)\biggr].
    \end{split}\end{equation} The Eve contribution wrong coincidence probability is \begin{equation}\begin{split}
    \mathrm{Pr}^\mathfrak{E}_\mathrm{w}(\theta)&=\frac{1}{2}\sum^2_{\substack{i=1 \\ i\neq j}}\mathrm{Pr}_{\mathrm{I:\mathcal{X}_{+,i}}}\biggl[ r \biggl(\mathrm{Pr}_{\mathrm{S:\mathcal{X}_{+,i}}\vert\mathrm{E:\mathcal{Y}_{+,i}}}(\theta_\mathrm{T})\mathrm{Pr}_{\mathrm{S:\mathcal{Y}_{+,i}}\vert\mathrm{I:\mathcal{X}_{+,i}}}(\theta_\mathrm{T})+\mathrm{Pr}_{\mathrm{S:\mathcal{X}_{+,i}}\vert\mathrm{E:\mathcal{Y}_{+,j}}}(\theta_\mathrm{T})\mathrm{Pr}_{\mathrm{S:\mathcal{Y}_{+,j}}\vert\mathrm{I:\mathcal{X}_{+,i}}}(\theta_\mathrm{T})\biggr)+ \\ &+(1-r)\biggl(\mathrm{Pr}_{\mathrm{S:\mathcal{X}_{+,i}}\vert\mathrm{E:\mathcal{Y}_{\times,i}}}(\theta_\mathrm{T})\mathrm{Pr}_{\mathrm{S:\mathcal{Y}_{\times,i}}\vert\mathrm{I:\mathcal{X}_{+,i}}}(\theta_\mathrm{T})+\mathrm{Pr}_{\mathrm{S:\mathcal{X}_{+,i}}\vert\mathrm{E:\mathcal{X}_{\times,j}}}(\theta_\mathrm{T})\mathrm{Pr}_{\mathrm{S:\mathcal{Y}_{\times,j}}\vert\mathrm{I:\mathcal{X}_{+,i}}}(\theta_\mathrm{T})\biggr)\biggr]+ \\ &+\frac{1}{2}\sum^2_{\substack{i=1 \\ i\neq j}}\mathrm{Pr}_{\mathrm{I:\mathcal{X}_{\times,i}}}\biggl[ r \biggl(\mathrm{Pr}_{\mathrm{S:\mathcal{X}_{\times,j}}\vert\mathrm{E:\mathcal{Y}_{+,i}}}(\theta_\mathrm{T})\mathrm{Pr}_{\mathrm{S:\mathcal{Y}_{+,i}}\vert\mathrm{I:\mathcal{X}_{\times,i}}}(\theta_\mathrm{T})+\mathrm{Pr}_{\mathrm{S:\mathcal{X}_{\times,j}}\vert\mathrm{E:\mathcal{Y}_{+,j}}}(\theta_\mathrm{T})\mathrm{Pr}_{\mathrm{S:\mathcal{Y}_{+,j}}\vert\mathrm{I:\mathcal{X}_{\times,i}}}(\theta_\mathrm{T})\biggr)+ \\ &+(1-r)\biggl(\mathrm{Pr}_{\mathrm{S:\mathcal{X}_{\times,j}}\vert\mathrm{E:\mathcal{Y}_{\times,i}}}(\theta_\mathrm{T})\mathrm{Pr}_{\mathrm{S:\mathcal{Y}_{\times,i}}\vert\mathrm{I:\mathcal{X}_{\times,i}}}(\theta_\mathrm{T})+\mathrm{Pr}_{\mathrm{S:\mathcal{X}_{\times,j}}\vert\mathrm{E:\mathcal{X}_{\times,j}}}(\theta_\mathrm{T})\mathrm{Pr}_{\mathrm{S:\mathcal{Y}_{\times,j}}\vert\mathrm{I:\mathcal{X}_{\times,i}}}(\theta_\mathrm{T})\biggr)\biggr]+ \\
    &+\frac{1}{2}\sum^2_{\substack{i=1 \\ i\neq j}}\biggl[\mathrm{Pr}_{\mathrm{I:\mathcal{X}_{+,i}}}\mathrm{vac}(\mathcal{X}_{+,i})\biggl(r\mathrm{Pr}_{\mathrm{S0:\mathcal{Y}_+\vert I:\mathcal{X}_{+,i}}}(\theta_\mathrm{T})+(1-r)\mathrm{Pr}_{\mathrm{S0:\mathcal{Y}_\times\vert I:\mathcal{X}_{+,i}}}(\theta_\mathrm{T})\biggr)+\\ &+\mathrm{Pr}_{\mathrm{I:\mathcal{X}_{\times,i}}}\mathrm{vac}(\mathcal{X}_{\times,j})\biggl(r\mathrm{Pr}_{\mathrm{S0:\mathcal{Y}_+\vert I:\mathcal{X}_{\times,i}}}(\theta_\mathrm{T})+(1-r)\mathrm{Pr}_{\mathrm{S0:\mathcal{Y}_\times\vert I:\mathcal{X}_{\times,i}}}(\theta_\mathrm{T})\biggr)\biggr]
    \end{split}\end{equation} The Eve contribution double coincidence probability is \begin{equation}\begin{split}
    \mathrm{Pr}^\mathfrak{E}_\mathrm{wc}(\theta)&=\frac{1}{2}\sum^2_{\substack{i=1 \\ i\neq j}}\mathrm{Pr}_{\mathrm{I:\mathcal{X}_{+,i}}}\biggl[ r \biggl(\mathrm{Pr}_{\mathrm{S:\mathcal{X}_{+,i},\mathcal{X}_{+,j}}\vert\mathrm{E:\mathcal{Y}_{+,i}}}(\theta_\mathrm{T})\mathrm{Pr}_{\mathrm{S:\mathcal{Y}_{+,i}}\vert\mathrm{I:\mathcal{X}_{+,i}}}(\theta_\mathrm{T})+\mathrm{Pr}_{\mathrm{S:\mathcal{X}_{+,j},\mathcal{X}_{+,i}}\vert\mathrm{E:\mathcal{Y}_{+,j}}}(\theta_\mathrm{T})\mathrm{Pr}_{\mathrm{S:\mathcal{Y}_{+,j}}\vert\mathrm{I:\mathcal{X}_{+,i}}}(\theta_\mathrm{T})\biggr)+ \\ &+(1-r)\biggl(\mathrm{Pr}_{\mathrm{S:\mathcal{X}_{+,j},\mathcal{X}_{+,i}}\vert\mathrm{E:\mathcal{Y}_{\times,i}}}(\theta_\mathrm{T})\mathrm{Pr}_{\mathrm{S:\mathcal{Y}_{\times,i}}\vert\mathrm{I:\mathcal{X}_{+,i}}}(\theta_\mathrm{T})+\mathrm{Pr}_{\mathrm{S:\mathcal{X}_{+,i}\mathcal{X}_{+,j}}\vert\mathrm{E:\mathcal{X}_{\times,j}}}(\theta_\mathrm{T})\mathrm{Pr}_{\mathrm{S:\mathcal{Y}_{\times,j}}\vert\mathrm{I:\mathcal{X}_{+,i}}}(\theta_\mathrm{T})\biggr)\biggr]+ \\ &+\frac{1}{2}\sum^2_{\substack{i=1 \\ i\neq j}}\mathrm{Pr}_{\mathrm{I:\mathcal{X}_{\times,i}}}\biggl[ r \biggl(\mathrm{Pr}_{\mathrm{S:\mathcal{X}_{\times,i},\mathcal{X}_{\times,j}}\vert\mathrm{E:\mathcal{Y}_{+,i}}}(\theta_\mathrm{T})\mathrm{Pr}_{\mathrm{S:\mathcal{Y}_{+,i}}\vert\mathrm{I:\mathcal{X}_{\times,i}}}(\theta_\mathrm{T})+\mathrm{Pr}_{\mathrm{S:\mathcal{X}_{\times,j},\mathcal{X}_{\times,i}}\vert\mathrm{E:\mathcal{Y}_{+,j}}}(\theta_\mathrm{T})\mathrm{Pr}_{\mathrm{S:\mathcal{Y}_{+,j}}\vert\mathrm{I:\mathcal{X}_{\times,i}}}(\theta_\mathrm{T})\biggr)+ \\ &+(1-r)\biggl(\mathrm{Pr}_{\mathrm{S:\mathcal{X}_{\times,i}\mathcal{X}_{\times,j}}\vert\mathrm{E:\mathcal{Y}_{\times,i}}}(\theta_\mathrm{T})\mathrm{Pr}_{\mathrm{S:\mathcal{Y}_{\times,i}}\vert\mathrm{I:\mathcal{X}_{\times,i}}}(\theta_\mathrm{T})+\mathrm{Pr}_{\mathrm{S:}\mathcal{X}_{\times,j},\mathrm{\mathcal{X}_{\times,i}}\vert\mathrm{E:\mathcal{X}_{\times,j}}}(\theta_\mathrm{T})\mathrm{Pr}_{\mathrm{S:\mathcal{Y}_{\times,j}}\vert\mathrm{I:\mathcal{X}_{\times,i}}}(\theta_\mathrm{T})\biggr)\biggr]+ \\
    &+\frac{1}{2}\sum^2_{\substack{i=1 \\ i\neq j}}\biggl[\mathrm{Pr}_{\mathrm{I:\mathcal{X}_{+,i}}}\mathrm{vac}(\mathcal{X}_{+,i},\mathcal{X}_{+,j})\biggl(r\mathrm{Pr}_{\mathrm{S0:\mathcal{Y}_+\vert I:\mathcal{X}_{+,i}}}(\theta_\mathrm{T})+(1-r)\mathrm{Pr}_{\mathrm{S0:\mathcal{Y}_\times\vert I:\mathcal{X}_{+,i}}}(\theta_\mathrm{T})\biggr)+\\ &+\mathrm{Pr}_{\mathrm{I:\mathcal{X}_{\times,i}}}\mathrm{vac}(\mathcal{X}_{\times,i},\mathcal{X}_{\times,j})\biggl(r\mathrm{Pr}_{\mathrm{S0:\mathcal{Y}_+\vert I:\mathcal{X}_{\times,i}}}(\theta_\mathrm{T})+(1-r)\mathrm{Pr}_{\mathrm{S0:\mathcal{Y}_\times\vert I:\mathcal{X}_{\times,i}}}(\theta_\mathrm{T})\biggr)\biggr]
    \end{split}\end{equation}
\end{widetext}
\subsection{Noise-only}
As mentioned earlier, we assume that both channels have the same background noise. This section details the click probabilities when there is no object present. I.e. the background noise click probabilities. The background noise contribution correct coincidence click probability is \begin{align}
    \mathrm{Pr}^\mathrm{B}_\mathrm{c}&=\frac{1}{2}\sum^2_{\substack{i=1 \\ i\neq j}}\mathrm{Pr}_{\mathrm{I:\mathcal{X}_{+,i}}}\mathrm{vac}(\mathcal{X}_{+,j})+\nonumber\\ &+\frac{1}{2}\sum^2_{i=1}\mathrm{Pr}_{\mathrm{I:\mathcal{X}_{\times,i}}}\mathrm{vac}(\mathcal{X}_{\times,i}).\end{align} The background noise contribution wrong coincidence click probability is \begin{align}
    \mathrm{Pr}^\mathrm{B}_\mathrm{w}&=\frac{1}{2}\sum^2_{i=1}\mathrm{Pr}_{\mathrm{I:\mathcal{X}_{+,i}}}\mathrm{vac}(\mathcal{X}_{+,i})+\nonumber\\ &+\frac{1}{2}\sum^2_{\substack{i=1 \\ i\neq j}}\mathrm{Pr}_{\mathrm{I:\mathcal{X}_{\times,i}}}\mathrm{vac}(\mathcal{X}_{\times,j}).\end{align} The background noise contribution double coincidence click probability is \begin{align}
    \mathrm{Pr}^\mathrm{B}_\mathrm{wc}&=\frac{1}{2}\sum^2_{\substack{i=1 \\ i\neq j}}\mathrm{Pr}_{\mathrm{I:\mathcal{X}_{+,i}}}\mathrm{vac}(\mathcal{X}_{+,j},\mathcal{X}_{+,i})+ \nonumber\\ &+\frac{1}{2}\sum^2_{\substack{i=1 \\ i\neq j}}\mathrm{Pr}_{\mathrm{I:\mathcal{X}_{\times,i}}}\mathrm{vac}(\mathcal{X}_{\times,i},\mathcal{X}_{\times,j}).\end{align}
\subsection{BB84-inspired system}
\label{appendix:weak_coherent_state_prob}
We compare our QI system with similar system which has a single-mode coherent state \begin{equation}\hat{\rho}_\mathrm{\alpha}=e^{-\vert \alpha \vert^2}\sum^\infty_{n=0}\frac{\vert \alpha\vert^{2n}}{n!}\vert n\rangle\langle n\vert\end{equation} as its light source, where we exclude off-diagonal terms due to the detectors we consider. The use of a weak coherent state is commonplace in a prepare-and-measure BB84 QKD system, therefore we refer to this QSI system as the BB84-inspired system. The architecture of this system is identical for the light Eve sends towards Alice and is similar to Fig.~\ref{fig:oursystem} sans the idler detector sub-system. This state of light has Poissonian photon statistics and we assume that the mean photon number $\bar{n}_\mathrm{\alpha}=\vert \alpha\vert^2$ is small enough to ensure that there are no multi-photon contributions, where $\alpha$ is the amplitude of the coherent state. Therefore, the probability of the BB84-inspired system light source with 2-photons is negligible \begin{equation}
    P(2)=e^{-\vert \alpha\vert^2}\frac{\vert \alpha\vert^4}{2}\approx 0.
\end{equation}We truncate the single-mode coherent state such that it only has vacuum and single-photon contributions, therefore we must renormalise with the factor $\mathcal{N}_\alpha=\mathrm{Tr}\bigl(e^{\vert \alpha\vert^2}\sum^1_{n=0}\frac{\vert \alpha\vert^{2n}}{n!}\vert n \rangle\langle n \vert \bigr)$. The resulting state for the BB84-inspired system is \begin{equation}
    \hat{\rho}_\alpha^{'}=c^\alpha_0\vert 0\rangle\langle 0\vert_\mathcal{Q}+c^\alpha_1\vert 1\rangle\langle 1\vert_\mathcal{Q},
\end{equation} where $c^\alpha_0=\frac{e^{\vert \alpha\vert^2}}{\mathcal{N}_\alpha}$ and $c^\alpha_1=\frac{e^{\vert \alpha\vert^2}\vert\alpha\vert^2}{\mathcal{N}_\alpha}$ and as this is a prepare-and-measure scheme we have selected $\mathcal{Q}$ as the mode Alice sends her photon in. Alice's truncated state in terms of the basis $(\mathcal{J},\mathcal{K})$ of the signal detectors (where once again $\theta_\mathrm{T}=0$ if it is for the untampered light pathway) is \begin{widetext}
\begin{equation}
    \hat{\rho}_\alpha^{'}=c^\alpha_0\vert 0\rangle\langle 0\vert_\mathcal{J}\otimes\vert 0\rangle\langle0\vert_\mathcal{K}+c^\alpha_1\big(\mathrm{cos}^2(\theta_\mathrm{T})\vert 1\rangle\langle 1\vert_\mathcal{J}\otimes\vert 0 \rangle\langle 0 \vert_\mathcal{K}+\mathrm{sin}^2(\theta_\mathrm{T})\vert 0\rangle\langle 0\vert_\mathcal{J}\otimes\vert 1\rangle\langle 1 \vert_\mathcal{K}\bigr).
\end{equation}This (unnormalised) state after the inclusion of noise and loss at the signal detector sub-system is \begin{align}
    \hat{\rho}^\mathrm{UNNORM}_{\mathrm{S:}\alpha}&=c_0^\alpha \hat{\rho}^\mathrm{B}_\mathcal{J}(\vert r_\mathcal{J}\vert^2 \bar{n}_\mathrm{B,\mathcal{J}})\otimes \hat{\rho}^\mathrm{B}_\mathcal{K}(\vert r_\mathcal{K}\vert^2 \bar{n}_\mathrm{B,\mathcal{K}})+ \nonumber\\ &+c^\alpha_1\biggl(\mathrm{cos}^2(\theta_\mathrm{T})\hat{\phi}(\bar{n}_{\mathrm{B}:\mathcal{J}},\vert t_\mathcal{J}\vert^2,\bar{n}_{\mathrm{B}:\mathcal{K}},\vert t_\mathcal{K}\vert^2)+\mathrm{sin}^2(\theta_\mathrm{T})\hat{\phi}(\bar{n}_{\mathrm{B}:\mathcal{K}},\vert t_\mathcal{K}\vert^2,\bar{n}_{\mathrm{B}:\mathcal{J}},\vert t_\mathcal{J}\vert^2)\biggr),\end{align} after which we normalise this state, which results in $\hat{\rho}_{\mathrm{S:}\alpha}=\hat{\rho}^\mathrm{UNNORM}_{\mathrm{S:}\alpha}/{\mathcal{N}_{\mathrm{S}:\alpha}}$, where $\mathcal{N}_{\mathrm{S}:\alpha}=\mathrm{Tr}\bigl(\hat{\rho}^\mathrm{UNNORM}_{\mathrm{S:}\alpha}\bigr)$. The following click probabilities apply for Eve or Alice's signal detectors (if it is untampered light incident at Alice's detectors). The click probability where the signal detector for mode $\mathcal{J}$ clicks and $\mathcal{K}$ does not is \begin{align}
    \mathrm{Pr}_{\mathrm{S:\mathcal{J}}\vert \alpha:\mathcal{Q}}&=\mathrm{Tr}\bigl(\hat{\rho}_{\mathrm{S:}\alpha}(\hat{1}_\mathcal{J}-\vert 0\rangle\langle 0 \vert_\mathcal{J})\otimes \vert 0\rangle\langle0\vert_\mathcal{K}\bigr), \nonumber \\ &=\frac{1}{\mathcal{N}_{\mathrm{S}:\alpha}}\biggl(c_0^\alpha\mathrm{vac}(\mathcal{J})+c_1^\alpha\bigl(\mathrm{cos}^2(\theta_\mathrm{T} )a_\mathrm{\mathcal{J}\vert 1}+\mathrm{sin}^2(\theta_\mathrm{T})a_\mathrm{\mathcal{K}\vert 0}\bigr)\biggr).
\end{align} The click probability where the signal detector for mode $\mathcal{K}$ clicks and $\mathcal{J}$ does not is \begin{align}
    \mathrm{Pr}_{\mathrm{S:\mathcal{K}}\vert \alpha:\mathcal{Q}}&=\mathrm{Tr}\bigl(\hat{\rho}_{\mathrm{S:}\alpha}(\hat{1}_\mathcal{K}-\vert 0\rangle\langle 0 \vert_\mathcal{K})\otimes \vert 0\rangle\langle0\vert_\mathcal{J}\bigr), \nonumber \\ &=\frac{1}{\mathcal{N}_{\mathrm{S}:\alpha}}\biggl(c_0^\alpha\mathrm{vac}(\mathcal{K})+c_1^\alpha\bigl(\mathrm{cos}^2(\theta_\mathrm{T} )a_\mathrm{\mathcal{J}\vert 0}+\mathrm{sin}^2(\theta_\mathrm{T})a_\mathrm{\mathcal{K}\vert 1}\bigr)\biggr).
\end{align} The click probability where the signal detector for both mode $\mathcal{J}$ and $\mathcal{K}$ clicks is \begin{align}
    \mathrm{Pr}_{\mathrm{S:\mathcal{J,K}}\vert \alpha:\mathcal{Q}}&=\mathrm{Tr}\bigl(\hat{\rho}_{\mathrm{S:}\alpha}(\hat{1}_\mathcal{J}-\vert 0\rangle\langle 0 \vert_\mathcal{J})\otimes (\hat{1}_\mathcal{K}-\vert 0\rangle\langle0\vert_\mathcal{K})\bigr), \nonumber \\ &=\frac{1}{\mathcal{N}_{\mathrm{S}:\alpha}}\biggl(c_0^\alpha\mathrm{vac}(\mathcal{J,K})+c_1^\alpha\bigl(\mathrm{cos}^2(\theta_\mathrm{T} )a_\mathrm{\mathcal{J,K}\vert 1}+\mathrm{sin}^2(\theta_\mathrm{T})a_\mathrm{\mathcal{K,J}\vert 1}\bigr)\biggr).
\end{align}
For click probability when neither of Eve's signal detectors for the modes in a set $\mathcal{L}\in \{\mathcal{J},\mathcal{K}\}$ clicks is \begin{align}
   \mathrm{Pr}_\mathrm{S0:\mathcal{L}\vert \alpha:\mathcal{Q}}(\theta_\mathrm{T})&= \mathrm{Tr}\left(\hat{\rho}_\mathrm{S:\alpha}\vert 0\rangle\langle 0\vert_{\mathcal{J}}\otimes\vert 0 \rangle\langle 0\vert_\mathcal{K}\right),\nonumber\\ &=\frac{1}{\mathcal{N}_{\mathrm{S:\alpha}}}\biggl(c^\alpha_0 \mathrm{vac}_0(\mathcal{J,K})+c^\alpha_1\bigl(q_2(\theta_\mathrm{T})a_{\mathcal{J,K}\vert 0}+q_1(\theta_\mathrm{T})a_{\mathcal{K,J}\vert 0}\bigr)\biggr).
\end{align}

The Alice's signal detector click probabilities when Eve's resent light is incident are the same for the BB84-inspired system as the QI system, as Alice's state of light does not affect the fact that Eve sends out a single photon in a mode instructed by her measurement. However, the click probabilities of contributions from Eve, Alice and noise will differ from the QI system. To note, the mode we prepare $\alpha:\mathcal{X}_{+/\times,\mathrm{i}}$ is correlated with the signal detector mode $\mathrm{S}:\mathcal{X}_\mathrm{+/\times,i}$, for example. Hence, starting with Alice (and omitting the $\theta_\mathrm{T}$ term), the Alice contribution correct coincidence click probability is 
\begin{equation}
    \mathrm{Pr}_\mathrm{c}^{\mathfrak{A}:\alpha}=\frac{1}{4}\sum^2_{i=1}( \mathrm{Pr}_{\mathrm{S:\mathcal{X}_{+,i}}\vert \alpha:\mathcal{X}_{+,i}}+\mathrm{Pr}_{\mathrm{S:\mathcal{X}_{\times,i}}\vert \alpha:\mathcal{X}_{\times,i}}). 
\end{equation} The Alice contribution wrong coincidence click probability is \begin{equation}
    \mathrm{Pr}_\mathrm{w}^{\mathfrak{A}:\alpha}=\frac{1}{4}\sum^2_{\substack{i=1 \\ i\neq j}}( \mathrm{Pr}_{\mathrm{S:\mathcal{X}_{+,j}}\vert \alpha:\mathcal{X}_{+,i}}+\mathrm{Pr}_{\mathrm{S:\mathcal{X}_{\times,j}}\vert \alpha:\mathcal{X}_{\times,i}}). 
\end{equation}Furthermore, the Alice contribution double coincidence click probability is \begin{equation}
    \mathrm{Pr}_\mathrm{wc}^{\mathfrak{A}:\alpha}=\frac{1}{4}\sum^2_{\substack{i=1 \\ i\neq j}}( \mathrm{Pr}_{\mathrm{S:\mathcal{X}_{+,i},\mathcal{X}_{+,j}}\vert \alpha:\mathcal{X}_{+,i}}+\mathrm{Pr}_{\mathrm{S:\mathcal{X}_{\times,i},\mathcal{X}_{\times,j}}\vert \alpha:\mathcal{X}_{\times,i}}). 
\end{equation}

We now move onto Eve, and the Eve contribution correct coincidence click probability is 
\begin{equation}
\begin{split}
    \mathrm{Pr}_\mathrm{c}^{\mathfrak{E}:\alpha}(\theta)&=\frac{1}{4}\sum^2_{\substack{i=1 \\ i\neq j}}\biggl(r\bigl(\mathrm{Pr}_\mathrm{S:\mathcal{X}_{+,i}\vert E:\mathcal{Y}_\mathrm{+,i}}(\theta_\mathrm{T})\mathrm{Pr}_\mathrm{S:\mathcal{Y}_{+,i}\vert \alpha:\mathcal{X}_\mathrm{+,i}}(\theta_\mathrm{T})+\mathrm{Pr}_\mathrm{S:\mathcal{X}_{+,i}\vert E:\mathcal{Y}_{+,\mathrm{j}}}(\theta_\mathrm{T})\mathrm{Pr}_\mathrm{S:\mathcal{Y}_{+,j}\vert \alpha:\mathcal{X}_\mathrm{+,i}}(\theta_\mathrm{T})\bigr)+ \\&+(1-r)\bigl(\mathrm{Pr}_\mathrm{S:\mathcal{X}_{+,i}\vert E:\mathcal{Y}_\mathrm{\times,i}}(\theta_\mathrm{T})\mathrm{Pr}_\mathrm{S:\mathcal{Y}_{\times,i}\vert \alpha:\mathcal{X}_\mathrm{+,i}}(\theta_\mathrm{T})+\mathrm{Pr}_\mathrm{S:\mathcal{X}_{+,i}\vert E:\mathcal{Y}_{\times,\mathrm{j}}}(\theta_\mathrm{T})\mathrm{Pr}_\mathrm{S:\mathcal{Y}_{\times,j}\vert \alpha:\mathcal{X}_\mathrm{+,i}}(\theta_\mathrm{T})\bigr)+ \\ &+r\bigl(\mathrm{Pr}_\mathrm{S:\mathcal{X}_{\times,i}\vert E:\mathcal{Y}_\mathrm{+,i}}(\theta_\mathrm{T})\mathrm{Pr}_\mathrm{S:\mathcal{Y}_{+,i}\vert \alpha:\mathcal{X}_\mathrm{\times,i}}(\theta_\mathrm{T})+\mathrm{Pr}_\mathrm{S:\mathcal{X}_{\times,i}\vert E:\mathcal{Y}_{+,\mathrm{j}}}(\theta_\mathrm{T})\mathrm{Pr}_\mathrm{S:\mathcal{Y}_{+,j}\vert \alpha:\mathcal{X}_\mathrm{\times,i}}(\theta_\mathrm{T})\bigr)+ \\&+(1-r)\bigl(\mathrm{Pr}_\mathrm{S:\mathcal{X}_{\times,i}\vert E:\mathcal{Y}_\mathrm{\times,i}}(\theta_\mathrm{T})\mathrm{Pr}_\mathrm{S:\mathcal{Y}_{\times,i}\vert \alpha:\mathcal{X}_\mathrm{\times,i}}(\theta_\mathrm{T})+\mathrm{Pr}_\mathrm{S:\mathcal{X}_{\times,i}\vert E:\mathcal{Y}_{\times,\mathrm{j}}}(\theta_\mathrm{T})\mathrm{Pr}_\mathrm{S:\mathcal{Y}_{\times,j}\vert \alpha:\mathcal{X}_\mathrm{\times,i}}(\theta_\mathrm{T})\bigr)+ \\&+\mathrm{vac}(\mathcal{X}_{+,i})\bigl(r\;\mathrm{Pr}_{\mathrm{S0:\mathcal{Y}_+}\vert \alpha:\mathcal{X}_{+,i}}(\theta_\mathrm{T})+(1-r)\;\mathrm{Pr}_{\mathrm{S0:\mathcal{Y}_\times}\vert \alpha:\mathcal{X}_{+,i}}(\theta_\mathrm{T})\bigr)+\\&+\mathrm{vac}(\mathcal{X}_{\times,i})\bigl(r\;\mathrm{Pr}_{\mathrm{S0:\mathcal{Y}_+}\vert \alpha:\mathcal{X}_{\times,i}}(\theta_\mathrm{T})+(1-r)\;\mathrm{Pr}_{\mathrm{S0:\mathcal{Y}_\times}\vert \alpha:\mathcal{X}_{\times,i}}(\theta_\mathrm{T})\bigr)\biggr).
    \end{split}
\end{equation}
The Eve contribution wrong coincidence click probability is  
\begin{equation}
\begin{split}
    \mathrm{Pr}_\mathrm{w}^{\mathfrak{E}:\alpha}(\theta)&=\frac{1}{4}\sum^2_{\substack{i=1 \\ i\neq j}}\biggl(r\bigl(\mathrm{Pr}_\mathrm{S:\mathcal{X}_{+,j}\vert E:\mathcal{Y}_\mathrm{+,i}}(\theta_\mathrm{T})\mathrm{Pr}_\mathrm{S:\mathcal{Y}_{+,i}\vert \alpha:\mathcal{X}_\mathrm{+,i}}(\theta_\mathrm{T})+\mathrm{Pr}_\mathrm{S:\mathcal{X}_{+,j}\vert E:\mathcal{Y}_{+,\mathrm{j}}}(\theta_\mathrm{T})\mathrm{Pr}_\mathrm{S:\mathcal{Y}_{+,j}\vert \alpha:\mathcal{X}_\mathrm{+,i}}(\theta_\mathrm{T})\bigr)+ \\&+(1-r)\bigl(\mathrm{Pr}_\mathrm{S:\mathcal{X}_{+,j}\vert E:\mathcal{Y}_\mathrm{\times,i}}(\theta_\mathrm{T})\mathrm{Pr}_\mathrm{S:\mathcal{Y}_{\times,i}\vert \alpha:\mathcal{X}_\mathrm{+,i}}(\theta_\mathrm{T})+\mathrm{Pr}_\mathrm{S:\mathcal{X}_{+,j}\vert E:\mathcal{Y}_{\times,\mathrm{j}}}(\theta_\mathrm{T})\mathrm{Pr}_\mathrm{S:\mathcal{Y}_{\times,j}\vert \alpha:\mathcal{X}_\mathrm{+,i}}(\theta_\mathrm{T})\bigr)+ \\ &+r\bigl(\mathrm{Pr}_\mathrm{S:\mathcal{X}_{\times,j}\vert E:\mathcal{Y}_\mathrm{+,i}}(\theta_\mathrm{T})\mathrm{Pr}_\mathrm{S:\mathcal{Y}_{+,i}\vert \alpha:\mathcal{X}_\mathrm{\times,i}}(\theta_\mathrm{T})+\mathrm{Pr}_\mathrm{S:\mathcal{X}_{\times,j}\vert E:\mathcal{Y}_{+,\mathrm{j}}}(\theta_\mathrm{T})\mathrm{Pr}_\mathrm{S:\mathcal{Y}_{+,j}\vert \alpha:\mathcal{X}_\mathrm{\times,i}}(\theta_\mathrm{T})\bigr)+ \\&+(1-r)\bigl(\mathrm{Pr}_\mathrm{S:\mathcal{X}_{\times,j}\vert E:\mathcal{Y}_\mathrm{\times,i}}(\theta_\mathrm{T})\mathrm{Pr}_\mathrm{S:\mathcal{Y}_{\times,i}\vert \alpha:\mathcal{X}_\mathrm{\times,i}}(\theta_\mathrm{T})+\mathrm{Pr}_\mathrm{S:\mathcal{X}_{\times,j}\vert E:\mathcal{Y}_{\times,\mathrm{j}}}(\theta_\mathrm{T})\mathrm{Pr}_\mathrm{S:\mathcal{Y}_{\times,j}\vert \alpha:\mathcal{X}_\mathrm{\times,i}}(\theta_\mathrm{T})\bigr)+ \\&+\mathrm{vac}(\mathcal{X}_{+,j})\bigl(r\;\mathrm{Pr}_{\mathrm{S0:\mathcal{Y}_+}\vert \alpha:\mathcal{X}_{+,i}}(\theta_\mathrm{T})+(1-r)\;\mathrm{Pr}_{\mathrm{S0:\mathcal{Y}_\times}\vert \alpha:\mathcal{X}_{+,i}}(\theta_\mathrm{T})\bigr)+\\&+\mathrm{vac}(\mathcal{X}_{\times,j})\bigl(r\;\mathrm{Pr}_{\mathrm{S0:\mathcal{Y}_+}\vert \alpha:\mathcal{X}_{\times,i}}(\theta_\mathrm{T})+(1-r)\;\mathrm{Pr}_{\mathrm{S0:\mathcal{Y}_\times}\vert \alpha:\mathcal{X}_{\times,i}}(\theta_\mathrm{T})\bigr)\biggr).
\end{split}
\end{equation}
The Eve contribution double coincidence click probability is 
\begin{equation}
\begin{split}
    \mathrm{Pr}_\mathrm{wc}^{\mathfrak{E}:\alpha}(\theta)&=\frac{1}{4}\sum^2_{\substack{i=1 \\ i\neq j}}\biggl(r\bigl(\mathrm{Pr}_\mathrm{S:\mathcal{X}_{+,i},\mathcal{X}_{+,j}\vert E:\mathcal{Y}_\mathrm{+,i}}(\theta_\mathrm{T})\mathrm{Pr}_\mathrm{S:\mathcal{Y}_{+,i}\vert \alpha:\mathcal{X}_\mathrm{+,i}}(\theta_\mathrm{T})+\mathrm{Pr}_\mathrm{S:\mathcal{X}_{+,j},\mathcal{X}_{+,i}\vert E:\mathcal{Y}_{+,\mathrm{j}}}(\theta_\mathrm{T})\mathrm{Pr}_\mathrm{S:\mathcal{Y}_{+,j}\vert \alpha:\mathcal{X}_\mathrm{+,i}}(\theta_\mathrm{T})\bigr)+ \\&+(1-r)\bigl(\mathrm{Pr}_\mathrm{S:\mathcal{X}_{+,j},\mathcal{X}_{+,i}\vert E:\mathcal{Y}_\mathrm{\times,i}}(\theta_\mathrm{T})\mathrm{Pr}_\mathrm{S:\mathcal{Y}_{\times,i}\vert \alpha:\mathcal{X}_\mathrm{+,i}}(\theta_\mathrm{T})+\mathrm{Pr}_\mathrm{S:\mathcal{X}_{+,i},\mathcal{X}_{+,j}\vert E:\mathcal{Y}_{\times,\mathrm{j}}}(\theta_\mathrm{T})\mathrm{Pr}_\mathrm{S:\mathcal{Y}_{\times,j}\vert \alpha:\mathcal{X}_\mathrm{+,i}}(\theta_\mathrm{T})\bigr)+ \\ &+r\bigl(\mathrm{Pr}_\mathrm{S:\mathcal{X}_{\times,i},\mathcal{X}_{\times,j}\vert E:\mathcal{Y}_\mathrm{+,i}}(\theta_\mathrm{T})\mathrm{Pr}_\mathrm{S:\mathcal{Y}_{+,i}\vert \alpha:\mathcal{X}_\mathrm{\times,i}}(\theta_\mathrm{T})+\mathrm{Pr}_\mathrm{S:\mathcal{X}_{\times,j},\mathcal{X}_{\times,i}\vert E:\mathcal{Y}_{+,\mathrm{j}}}(\theta_\mathrm{T})\mathrm{Pr}_\mathrm{S:\mathcal{Y}_{+,j}\vert \alpha:\mathcal{X}_\mathrm{\times,i}}(\theta_\mathrm{T})\bigr)+ \\&+(1-r)\bigl(\mathrm{Pr}_\mathrm{S:\mathcal{X}_{\times,i},\mathcal{X}_{\times,j}\vert E:\mathcal{Y}_\mathrm{\times,i}}(\theta_\mathrm{T})\mathrm{Pr}_\mathrm{S:\mathcal{Y}_{\times,i}\vert \alpha:\mathcal{X}_\mathrm{\times,i}}(\theta_\mathrm{T})+\mathrm{Pr}_\mathrm{S:\mathcal{X}_{\times,j},\mathcal{X}_{\times,i}\vert E:\mathcal{Y}_{\times,\mathrm{j}}}(\theta_\mathrm{T})\mathrm{Pr}_\mathrm{S:\mathcal{Y}_{\times,j}\vert \alpha:\mathcal{X}_\mathrm{\times,i}}(\theta_\mathrm{T})\bigr)+ \\&+\mathrm{vac}(\mathcal{X}_{+,i},\mathcal{X}_{+,j})\bigl(r\;\mathrm{Pr}_{\mathrm{S0:\mathcal{Y}_+}\vert \alpha:\mathcal{X}_{+,i}}(\theta_\mathrm{T})+(1-r)\;\mathrm{Pr}_{\mathrm{S0:\mathcal{Y}_\times}\vert \alpha:\mathcal{X}_{+,i}}(\theta_\mathrm{T})\bigr)+\\&+\mathrm{vac}(\mathcal{X}_{\times,i},\mathcal{X}_{\times,j})\bigl(r\;\mathrm{Pr}_{\mathrm{S0:\mathcal{Y}_+}\vert \alpha:\mathcal{X}_{\times,i}}(\theta_\mathrm{T})+(1-r)\;\mathrm{Pr}_{\mathrm{S0:\mathcal{Y}_\times}\vert \alpha:\mathcal{X}_{\times,i}}(\theta_\mathrm{T})\bigr)\biggr).
\end{split}
\end{equation}
\end{widetext}

The background noise contribution for BB84-inspired differs from QI also. The background noise correct coincidence click probability is equal to the wrong coincidence click probability and it is \begin{equation}
    \mathrm{Pr}^{\mathrm{B}:\alpha}_\mathrm{c}=\mathrm{Pr}^{\mathrm{B}:\alpha}_\mathrm{w}=\frac{1}{4}\sum^2_{i=1}\biggl(\mathrm{vac}(\mathcal{X}_{+,i})+\mathrm{vac}(\mathcal{X}_{\times,i})\biggr).
\end{equation}The background noise double coincidence click probability is 
\begin{align}
   \mathrm{Pr}^{\mathrm{B}:\alpha}_\mathrm{wc}&=\frac{1}{4}\sum^2_{\substack{i=1 \\ i\neq j}}\biggl(\mathrm{vac}(\mathcal{X}_{+,i},\mathcal{X}_{+,j})+ \nonumber \\&+\mathrm{vac}(\mathcal{X}_{\times,i},\mathcal{X}_{\times,j})\biggr).
    \end{align}

The real and false channel for the BB84 system is defined as the QI system is, albeit with an appropriate substitution of the Alice/Eve/noise coincidence probabilities.
\section{Proof}
\label{appendix:proof}
If Alice measures that both $\hat{\mathrm{Pr}}^\mathfrak{R/F}_\mathrm{w}-\mathrm{Pr}^\mathrm{B}_\mathrm{w}>0$, then she chooses the channel with the larger k-factor to be the one which contains her own light $k_\mathfrak{R}>k_\mathfrak{F}$. By definition, Alice's measurement of the false channel infers that $\mathrm{Pr}^\mathfrak{E}_\mathrm{c}-\mathrm{Pr}^\mathrm{B}_\mathrm{c}>\mathrm{Pr}^\mathfrak{E}_\mathrm{w}-\mathrm{Pr}^\mathrm{B}_\mathrm{w}>0$.

We prove that $k_\mathfrak{R}>k_\mathfrak{F}$ is true by contradiction. Assume that $k_\mathfrak{F}>k_\mathfrak{R}$, using the k-factor definitions we can restate this as $k_\mathfrak{F}>k_\mathfrak{F}+g$, where $g$ is a constant defined as \begin{widetext} \begin{equation}
    g=\frac{(\mathrm{Pr}^\mathfrak{A}_\mathrm{c}-\mathrm{Pr}^\mathrm{B}_\mathrm{c})(\mathrm{Pr}^\mathfrak{E}_\mathrm{w}-\mathrm{Pr}^\mathrm{B}_\mathrm{w})-(\mathrm{Pr}^\mathfrak{E}_\mathrm{c}-\mathrm{Pr}^\mathrm{B}_\mathrm{c})(\mathrm{Pr}^\mathfrak{A}_\mathrm{w}-\mathrm{Pr}^\mathrm{B}_\mathrm{w})}{(\mathrm{Pr}^\mathfrak{E}_\mathrm{w}-\mathrm{Pr}^\mathrm{B}_\mathrm{w})\biggl((\mathrm{Pr}^\mathfrak{A}_\mathrm{w}-\mathrm{Pr}^\mathrm{B}_\mathrm{w})+\frac{p_\mathfrak{R}}{1-p}(\mathrm{Pr}^\mathfrak{E}_\mathrm{w}-\mathrm{Pr}^\mathrm{B}_\mathrm{w})\biggr)},
\end{equation}
\end{widetext} we can prove by contradiction as $k_\mathfrak{F}\not>k_\mathfrak{F}+g$ if $g>0$, which can be done by showing that the numerator and denominator of $g$ are both positive. If we show that $\mathrm{Pr}^\mathfrak{A}_\mathrm{c}(0<\xi<1)-\mathrm{Pr}^\mathrm{B}_\mathrm{c}>0$, then the numerator of $g$ is guaranteed to be positive as $\mathrm{Pr}^\mathfrak{A}_\mathrm{w}-\mathrm{Pr}^\mathrm{B}_\mathrm{w}\leq 0$. The denominator of $g$ equals the product of the denominators of $k_\mathfrak{F}$ and  $k_\mathfrak{R}$, and if $\mathrm{Pr}^\mathfrak{A}_\mathrm{c}(0<\xi<1)-\mathrm{Pr}^\mathrm{B}_\mathrm{c}>0$ was shown, then the denominator of $g$ must be positive, as $k_\mathfrak{R}>0$ and $k_\mathfrak{F}>0$.
\section{Monte-Carlo}
\label{appendix:monte_carlo}
We have a M.C. simulation for the system we describe. This numerical approach means we can find the covariance of the noise contribution and the object present distribution which allows us to analyse the noise-reduced distributions. Instead of relying upon analysis with the expectation values only. 

The number of idler clicks is a Poisson random variable dictated by the idler click probability. The rest of the random variables in our simulation are conditionally dependent upon the number of idler clicks. If there has been enough simulation runs our resultant statistics matches with the analytic approach for the noise-reduced distributions to use the expected number of idler clicks. We do not consider the probability of interception $p$ or probability of $p_\mathfrak{F/R}$ false/real channel interception as a random variable, as the dependence between the real and false channel is negligible when the background noise is much greater than the signal (from Alice and Eve) contribution $\bar{n}\ll \bar{n}_\mathrm{B}$. We simulate the dependence of the random variable for the noise distribution and the object present distribution by using the same seed random number when sampling for both of these random variables. It is from this simulated dependence we can calculate the covariance. However, in practice, the clicks that originate from Alice/Eve or noise cannot be distinguished and so this dependence can only be calculated via our M.C. simulation.

\bibliographystyle{apsrev4-2}
\bibliography{export}
\end{document}